\documentclass[12pt]{article}
\usepackage{fullpage,epsfig,graphics,amsbsy,amssymb,cancel,bbm}
\usepackage{psfrag,hyperref,color,slashed}
\usepackage{graphicx,verbatim}
\usepackage{amsfonts,mathdots,dsfont}
\usepackage{amssymb,amsmath,amscd,mathrsfs}
\usepackage{mathabx}
\usepackage{tikz}
\usetikzlibrary{arrows,chains,matrix,positioning,scopes,snakes}

\makeatletter
\tikzset{join/.code=\tikzset{after node path={%
\ifx\tikzchainprevious\pgfutil@empty\else(\tikzchainprevious)%
edge[every join]#1(\tikzchaincurrent)\fi}}}
\makeatother

\tikzset{>=stealth',every on chain/.append style={join},
         every join/.style={->}}
\tikzset{
    >=stealth',
    punkt/.style={
           rectangle,
           rounded corners,
           draw=black, very thick,
           text width=6.5em,
           minimum height=2em,
           text centered},
    pil/.style={
           ->,
           thick,
           shorten <=2pt,
           shorten >=2pt,}
}

\newcommand{\BB}{\mathbb}

\newcommand{\FR}{\mathfrak}

\newcommand{\be}{\begin{equation}}
\newcommand{\ee}{\end{equation}}
\newcommand{\bea}{\begin{eqnarray}}
\newcommand{\eea}{\end{eqnarray}}
\newcommand{\nn}{\nonumber}
\newcommand{\Tr}{\textrm{Tr}}

\newcommand{\im}{\textrm{Im}\,}

\newcommand{\reeb}{\textrm{\scriptsize{$R$}}}
\newcommand{\sreeb}{\textrm{\tiny{$R$}}}

\def\gb{\beta}

\def\gd{\delta}
\def\ep{\epsilon}

\def\gs{\sigma}

\def\go{\omega}

\DeclareMathAlphabet{\mathpzc}{OT1}{pzc}{m}{it}

\newcommand{\qed}{\nobreak \ifvmode \relax \else
      \ifdim\lastskip<1.5em \hskip-\lastskip
      \hskip1.5em plus0em minus0.5em \fi \nobreak
      \vrule height0.5em width0.5em depth0.00em\fi}

\begin{document}
\renewcommand{\theequation}{\thesection.\arabic{equation}}
\setcounter{page}{0}

\thispagestyle{empty}
\begin{flushright} \small
UUITP-20/15\\
 \end{flushright}
\smallskip
\begin{center} \LARGE
{\bf  Modular properties of full 5D SYM partition function}
 \\[12mm] \normalsize
{\bf  Jian Qiu$^{a,b}$, Luigi Tizzano$^b$, Jacob Winding$^b$ and Maxim Zabzine$^b$} \\[8mm]
 {\small\it
${}^a$Mathematics Institute, Uppsala University\\
     Box 480, SE-75106 Uppsala, Sweden\\
      \vspace{.5cm}
${}^b$Department of Physics and Astronomy,
     Uppsala university,\\
     Box 516,
     SE-75120 Uppsala,
     Sweden\\
   }
\end{center}
\vspace{7mm}
\begin{abstract}
 \noindent
We study properties of the full partition function for the $U(1)$ 5D $\mathcal{N}=2^*$ gauge theory with adjoint hypermultiplet of mass $M$.
This theory is ultimately related to abelian 6D (2,0) theory.  We construct the full non-perturbative
  partition function on toric Sasaki-Einstein manifolds by gluing flat copies of the Nekrasov partition function and we
   express the full partition function in terms of the generalized double elliptic gamma function $G_2^C$ associated
    with a certain moment map cone $C$.
   The answer exhibits a curious $SL(4,\BB{Z})$ modular property. Finally, we propose a set of rules to construct the partition function that resembles the calculation of 5d supersymmetric partition function with the insertion of defects of various co-dimensions.
\end{abstract}

\eject
\normalsize
\tableofcontents

\section{Introduction}

The 5D supersymmetric gauge theories provide  a nice testing ground for the study of higher dimensional gauge theories.
 By themselves the 5D supersymmetric Yang-Mills theories are not renormalizable, however one can study their
  UV completions. Some of the 5D supersymmetric Yang-Mills theories are closely related to $\mathcal{N}=1$ superconformal
   field theories which are believed to be consistent quantum field theories. The $\mathcal{N}=2$ 5D supersymmetric Yang-Mills theory
    is unique and it is believed that its UV completion is the superconformal $(2,0)$ theory in six dimensions.
 By compactifying the $(2,0)$ theory on a circle of radius $R_6$, the theory reduces to the five dimensional
maximally supersymmetric Yang-Mills theory \cite{Witten:1995zh, Seiberg:1997ax}. In such case the coupling constant can be expressed as
\be
R_6 = \frac{g_{YM}^2}{4\pi}\;,
\ee
this relation follows from the identification of the Kaluza-Klein modes in the $(2,0)$ theory
with the instanton particles in the five dimensional theory as was suggested in \cite{Douglas:2010iu, Lambert:2010iw}.
 Non-perturbative effects of the 5d theory provide nontrivial information about the 6d theory compactified on the circle.
  In this paper we study the $\mathcal{N}=2^*$  5D supersymmetric Yang-Mills theory which corresponds to a vector multiplet plus
    a hypermultiplet in the adjoint representation with arbitrary mass $M$. We concentrate on the abelian version of the $\mathcal{N}=2^*$ on toric Sasaki-Einstein manifolds.  \\

The simplest example of toric Sasaki-Einstein manifold is $S^5$.   The partition function on $S^5$
 can be calculated via the supersymmetric localization technique pioneered in \cite{Pestun:2007rz} and the result
  was obtained in \cite{Kallen:2012cs,Kim:2012ava,Kallen:2012va}.
In  \cite{Kim:2012ava, Kim:2012qf}  the calculation of the partition function on $S^5$  for $\mathcal{N}=2^*$  5D gauge theory  has been
 discussed in the context of the $(2,0)$ theory on $S^5 \times S^1$.  It is believed that the partition function on
 $S^5$  for $\mathcal{N}=2^*$  5D gauge theory corresponds to a supersymmetric index counting BPS states of the $(2,0)$ theory on $\mathbb{C}^3$ in the radial quantization. The corresponding supersymmetric index  is   known as the superconformal index, for
  six dimensional superconformal theories it was defined in \cite{Bhattacharya:2008zy}.   \\

The localization calculation for the five-sphere has been used afterwards for a variety of purposes. For instance, in the context of AdS/CFT this has been used to obtain the $N^3$ behavior of the free energy of the (2,0) theory \cite{Minahan:2013jwa}. Another interesting development has been the study of the structure of the 5d partition functions through the holomorphic blocks as in \cite{Nieri:2013yra,Nieri:2013vba}. Finally, the authors of \cite{Lockhart:2012vp} proposed a definition of the non-perturbative topological string through the analysis of the five-sphere partition functions.\\

The study of the 5d partition function on other curved backgrounds was initiated in \cite{Qiu:2013pta}, in which the partition function for a family of five dimensional Sasaki-Einstein manifolds denoted $Y^{p,q}$ was calculated. Further calculations \cite{Qiu:2013aga} show that the answer can be factorised in certain building blocks extending the results in \cite{Nieri:2013yra,Nieri:2013vba}. Actually, $Y^{p,q}$ is just an example of a larger class of five dimensional manifolds known as toric Sasaki-Einstein (SE) manifolds. It is possible to define and calculate the partition function of a 5d theory on any toric SE manifold. This was done in \cite{Qiu:2014oqa} and also in that case the answer has a factorised form extending all the previously results on $S^5$ and $Y^{p,q}$. Let us stress that all these results about factorisation are only checked for the perturbative sector and on simply connected manifolds. As for the instanton sector, in the literature the factorisation is taken as the definition and a first principle computation is absent.\\

Let us outline schematically the main idea.
The building blocks mentioned above are associated with the so called closed Reeb orbits in the contact manifold $X$. In a neighborhood of such orbits, the geometry looks like $\mathbb{C}^2\times_{\ep} S^1$, where $\times_{\ep}$ means that one imposes a twisted periodic boundary condition along the $S^1$. The twisting is given by two $U(1)$'s acting on the two factors of $\BB{C}$ with two equivariant parameters $\ep,\ep'$. In this notation, the factorisation has the structure (more precise formula will come later)
\bea
Z^{\mathrm{Pert}}_X=\prod_i Z^{\mathrm{Pert}}_{\mathbb{C}^2\times_{\ep} S^1}(\gb_i,\ep_i,\ep_i')~,\nn
\eea
where $Z^{\mathrm{pert}}_{\mathbb{C}^2\times_{\ep} S^1}$ is the perturbative part of the Nekrasov partition function computed on $\mathbb{C}^2\times_{\ep} S^1$, and $\gb_i$ is the radius of the $S^1$.
We denote by $\gs\in\FR{h}$ the Coulomb branch parameter i.e. the weight of the action by the maximal torus of the gauge group, we can take as a working definition for the perturbative part of the Nekrasov partition function the following infinite product
\bea
 \prod_{p,q=0}^{\infty}\big(1-e^{2\pi i\gb(\gs+p\ep+q\ep')}\big)~,\label{crude}
\eea
but $\ep,\ep'$ must be given a small imaginary part for the product to converge, a more careful definition is given in \eqref{qfac}.
The next step is based on a crucial observation by \cite{Lockhart:2012vp} where the authors express the perturbative part as a special function known as triple sine function and then use the known factorisation property proven in \cite{MR2101221}. At the same time the perturbative part for general toric SE manifolds is expressed as a newly constructed generalized triple sine function, which
can be written as a product within a lattice
\bea S^{C_{\mu}}_3(z|\vec\go)=\prod_{\vec n\in\BB{Z}^3\cap C_{\mu}}\big(z+\vec n\cdotp\vec \go\big)\prod_{\vec n\in\BB{Z}^3\cap C^{\circ}_{\mu}}\big(\vec n\cdotp\vec \go-z\big)~,\nn\eea
where $C_{\mu}$ is a cone in $\BB{R}^3$, which is the image of the moment map of the torus action on the manifold, and $C_{\mu}^{\circ}$ is the interior of $C_{\mu}$.
This infinite product can be regulated using Riemann zeta function, provided that  the real part of $\vec\go$ is within the dual cone $C_{\mu}^{\vee}$. Then one can prove that $S^{C_{\mu}}_3$ factorises similarly into perturbative Nekrasov partition functions. Notice that, at least in the perturbative case,  this factorisation follows roughly from the localisation property of certain differential operators, even though in this way, one misses some important Bernoulli factors.\\

The goal of this work is to extend the observed factorisation property of the partition function beyond the perturbative case. We limit ourselves to the case of a $U(1)$ theory with an adjoint hypermultiplet of the mass $M$.
 The case with non-abelian  theory is complicated and we present  some short speculative comments
 in the conclusion. The full partition function in the abelian case can be calculated explicitly and recasted as the ratio of two double elliptic gamma functions. For the five-sphere case this observation was made in \cite{Lockhart:2012vp}, here we can extend their result for all toric SE manifolds. We can write our result in a concise way as
\bea
Z^{\mathrm{Full}}_X=\prod_i Z^{\mathrm{Full}}_{\mathbb{C}^2\times_{\ep} S^1}(\gb_i,\ep_i,\ep_i')~,\nn
\eea
with the same notation used before. This result has been obtained through the introduction of another new special function called the generalized double elliptic gamma function. The factorisation property for this new function as well as other useful properties that are relevant for the present paper have been proved
 in \cite{Tizzano:2014roa}. The generalized double elliptic gamma function can be written as
\bea
	G_2^{C_{\mu}} ( z | \vec \omega ) = \prod_{\vec n \in C\cap \mathbb{Z}^3 } ( 1 - e^{2\pi i ( z + \vec n \cdot \vec \omega ) } ) \prod_{\vec n \in C^\circ \cap \mathbb{Z}^3 } ( 1 - e^{2\pi i ( -z + \vec n \cdot \vec \omega ) } ) ~,
\eea
 while a more precise definition is given in \eqref{mult_ellip_gamma}.
This formula can be interpreted as product of ordinary double elliptic gamma functions associated to a subdvision of the cone. In this way, also for the case of the complete partition function we reach a factorised form expressed in terms of simple elementary blocks associated to the geometry of the cone.  The fact that we can write the final answer in terms of generalized double gamma function
is consistent with the 6D interpretation. Namely the generalized double elliptic gamma function can be written as
  infinite product
\be
G_2^{C_{\mu}} \sim \prod_{k = -\infty}^{\infty} S_3^{C_\mu} (k + z|\vec{\omega})~,
\ee
where the product over $k$ corresponds to the infinite tower of the Kaluza-Klein modes coming from the reduction on a circle and the identification of such modes with instantons.\\

Finally, we present another way to construct the full partition function which we interpret as a way to obtain a five dimensional toric SE manifold by gluing together a number of five-spheres after some appropriate surgeries. This was observed experimentally using the properties of
the special functions. However the construction is also motivated by geometry and physics considerations.
Gluing five-spheres together requires that we accounted for the degrees of freedom living on the co-dimension 2 and 4 locus where the gluing happens, indeed our rules for the construction involve some factors which look like supersymmetric indices for lower dimensional theories. This is very familiar in the context of supersymmetric defects, we will comment about some similarities with the structure found in the recent paper \cite{Bullimore:2014upa}.\\

The paper is organized as follows: in section \ref{s-flat-N} we review the explicit form of the Nekrasov partition function
for  abelian $\mathcal{N}=2^*$ theory on $\mathbb{C}^2 \times S^1$.  Section \ref{sec:ToricSE} presents the
construction of the full non-perturbative partition function for the same theory on any toric Sasaki-Einstein manifold.
In section \ref{sec:conemodularity} we study the properties of this partition function and
suggest a set of rules to construct the answer, that suggests the geometrical interpretation mentioned above. In section \ref{s-summary}
we give a summary of our results and list some open problems.  We supplement the paper by
two appendices: appendix \ref{app:special} contains the summary of the properties of the special functions and
appendix \ref{app:rewriting} collects the explicit formulas for the Nekrasov partition function
for  abelian $\mathcal{N}=2^*$ theory on $\mathbb{C}^2 \times S^1$.

\section{The 5D full abelian partition function on $\BB{C}^2\times_{\ep} S^1$}\label{s-flat-N}

The main subject of this paper is the five-dimensional $U(1)$ $\mathcal{N} = 2^*$ theory, i.e. the theory with a $U(1)$ vector multiplet and one massive adjoint hypermultiplet of mass $M$. At a perturbative level the theory is free, i.e. there are no interactions at all and the perturbative partition function is reduced to a simple Gaussian matrix model. Nevertheless, there is non-trivial information in the non-perturbative sector, where we see the appearance of small instantons in the quantum dynamics of the theory that correct the behavior of the partition function. These small instantons have a stringy origin \cite{Witten:1995gx} and they are crucial for, say, the $S^5$ case, where there are no smooth $U(1)$ instantons, thus these small instantons provide a UV completion for the 5d theory.  \\

As explained in \cite{Carlsson:2013jka} the 5D $\mathcal{N} = 2^*$ theory can be obtained through a compactification of the 6d $(2,0)$ theory on an elliptic curve with twisted boundary conditions. Or from the $M$5 brane point of view, one imposes twists on both the worldvolume of the $M$5 brane and the directions transverse to it.
The worldvolume then has four noncompact dimensions living in the Euclidean space $\mathbb{R}^4 \approx \mathbb{C}^2$, where the coordinates $1234$ are twisted by the parameters $(q_1,q_2)$, while the transverse $\mathbb{R}^5 \approx \mathbb{C}^2 \oplus \mathbb{R}^1$ has coordinates $56789$ twisted by
the parameters $(m/\sqrt{q_1q_2}, m^{-1}/\sqrt{q_1q_2}) \oplus 1$. The parameters $q_1,q_2,m$ are eventually written in terms of the equivariant parameters $\ep,\,\ep'$ and the mass $M$ of the gauge theory.\\

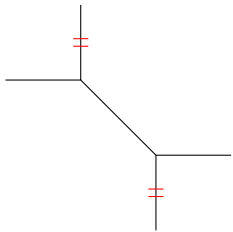
\begin{figure}[h]
\begin{center}
\begin{tikzpicture}[scale=1]
\draw [-,black] (-1,0) -- (0,0) -- (0, 1) ;
\draw[-,black] (0,0) -- (1,-1);
\draw[-,black] (1,-2) -- (1, -1) -- (2,-1);
\draw[-,red] (-.1, .45) -- (.1, .45);
\draw[-,red] (-.1, .55) -- (.1, .55);
\draw[-,red] (1-.1, .45-2) -- (1.1, .45-2);
\draw[-,red] (1-.1, .55-2) -- (1.1, .55-2);
\end{tikzpicture} \ \ \ \ \
\end{center}
\caption{The toric digram which realizes the $U(1)$ theory with adjoint mass.} \label{fig:adconifold}
\end{figure}

The partition function for this theory can be computed with various techniques. In the language of geometric engineering, the theory is associated to a toric diagram where two external legs has been identified along one direction. The toric diagram for this theory (shown in figure \ref{fig:adconifold}) was introduced in \cite{Hollowood:2003cv},
and the topological string partition function for the theory was computed in \cite{Iqbal:2008ra}. From the gauge theory point of view, the partition function for this theory on $\mathbb{C}^2 \times S^1$ was described in \cite{Nekrasov:2003rj,Nekrasov:2008kza} and
computed via equivariant localization in \cite{Poghossian:2008ge}. A mathematical proof of these results was recently established in \cite{Carlsson:2013jka}. The instanton part of the partition function was found to be
\be\label{eq:zinst}
	Z^{\mathrm{inst}}_{\mathbb{C}^2\times S^1} ( m, Q, q_1,q_2) = \exp \left [ \sum_{n=1}^\infty \frac{Q^n}{n m^n } \frac{ ( m^n - q_1^n ) (m^n - q_2^n ) } { (1-Q^n)(1-q_1^n)(1-q_2^n)} \right ]~,
\ee
with $q_1 = e^{2\pi i \epsilon},q_2 = e^{2\pi i \epsilon'}$, $\epsilon,\epsilon'$ being the equivariant rotation parameters of $\mathbb{C}^2$ mentioned previously. The parameter $m$ is related to the physical mass of the adjoint $M$ through $m = e^{2\pi i M}$. As is usual with instanton partition functions the quantity $Q=e^{- 2\pi i \beta}$, where $\beta$ is the radius of the $S^1$, plays the role of the instanton counting parameter.
The expression (\ref{eq:zinst}) must be supplemented by a perturbative contribution, which is $Q$-independent.
The perturbative partition function is given by\footnote{This expression differs by an overall sign flip in the numerator with respect to the one in \cite{Carlsson:2013jka}.}
\be
	Z^{\mathrm{pert}}_{\mathbb{C}^2\times S^1} = \exp \left [ \sum_{n=1}^\infty \frac{(q_1 q_2)^n }{n} \frac{m^n-1}{(1-q_1^n)(1-q_2^n)} \right ]~.
\ee
The full partition function $Z^{\mathrm{full} }_{\mathbb{C}^2 \times S^1} =Z^{\mathrm{pert}}Z^{\mathrm{inst}}$ is the same as the index
\be\label{eq:trh}
	 {\Tr}_{\mathcal{H}} \left( (-1)^{F} q_{1}^{J_{12} - R_1} q_{2}^{J_{34}- R_2} Q^{J_{56}- R_1} m^{R_2 - R_1} \right)~,
\ee
where $J_{ij}$ denote the rotation generators of $SO(6)$ and $R_1$ and $R_2$ denote the two Cartans of $Sp(4)$, the $R$-symmetry group, in an orthogonal basis. Equation (\ref{eq:trh}) represents the superconformal index of the free $(2,0)$ six dimensional theory computed as a trace over the Hilbert space $\mathcal{H}$ obtained by quantization on $\mathbb{C}^2 \times_{\ep} S^1$. 
\\

The full partition function can be written in terms of infinite products, or in terms of multiple elliptic gamma functions and multiple q-factorials. We describe this rewriting and other details about special functions in appendices \ref{app:special} and \ref{app:rewriting}. We can write
\be \label{eq:ZfullC2S1}
	Z^{\mathrm{full} }_{\mathbb{C}^2\times S^1} \sim \frac { G_2'  ( 0 | \epsilon, \epsilon',-\beta)}{G_2 ( M | \epsilon,\epsilon',-\beta)} ~,
\ee
 where we dropped certain pre-factors compared to \eqref{eq:zinst}, which consist of a product of $\eta$-functions. This is because the partition function suffers from some inherent ambiguity (which at the moment we do not know how
 to fix from the first principles\footnote{A possible way of fixing them is a comparison with the recent construction of $\mathcal{N}=2^*$ theories from string theory in \cite{Florakis:2015ied}, where the perturbative calculation is well defined.}), while dropping or keeping
 these additional $\eta$-functions does not affect the message of our paper.

In this formula the numerator corresponds to the contribution of the vector multiplet while the denominator is the contribution of the hypermultiplet. The numerator has a zero mode that needs to be removed or regularized, which is what we mean by writing the prime on $G_2$.
For technical reasons related to the presentation of the factorisation formulae used here, instead of removing the zero mode directly, we introduce a parameter $\delta$ to regulate the zero mode and write
\bea
G_2'  ( 0 | \epsilon, \epsilon',-\beta)\to G_2  ( \gb\gd | \epsilon, \epsilon',-\beta)~.\nn
\eea
Formula (\ref{eq:ZfullC2S1}) was already presented in \cite{Lockhart:2012vp}, and up to some ambiguities concerning the prefactor of $\eta$-functions, which in our notation is the q-factorial, we find the same answer.

\section{Full abelian partition function on toric Sasaki-Einstein manifolds}\label{sec:ToricSE}
\subsection{Geometry of the cone}
In order to discuss the abelian partition function on toric Sasaki-Einstein manifolds we need to introduce some well known facts about these manifolds, for more detailed explanations the reader may consult \cite{BoyerGalicki,2010arXiv1004.2461S}.\\

 Let $X$ be a 5-manifold, the metric cone of $X$ is given by $C(X)=X\times \mathbb{R}_{\gneq 0}$, with metric $G = d\mathfrak{r}^2 + \mathfrak{r}^2 g_X$, $\mathfrak{r}$ being the coordinate of the $\BB{R}$ factor. If $C(X)$ is K\"ahler, then $X$ is called \emph{Sasaki}, and if $C(X)$ is also Calabi-Yau, then $X$ is \emph{Sasaki-Einstein} (SE).
A Sasaki manifold has in particular a so called K-contact structure, with a Reeb vector $\reeb$ which is obtained by applying the complex structure to $\FR{r}\partial_{\FR{r}}$, the vector that scales $\FR{r}$. On $X$ there exists a transverse K\"ahler structure transverse to $\reeb$ which allows one to define a transverse Dolbeault operator, denoted $\bar\partial_H$. The orbits of $\reeb$ are generically not closed with the exception of a few isolated closed ones. In the neighbourhood of such closed orbit the geometry looks like $\BB{C}^2\times_{\ep}S^1$, with $\reeb$ along the $S^1$ direction.
If there is an effective, holomorphic and Hamiltonian action of $T^3$ on $C(X)$, such that the Reeb vector is given by a linear combination of the torus actions, then $X$ is called \emph{toric}.\\

 Let $\vec \mu$ be the moment map of the three torus actions.
The image of $\vec \mu$, i.e. $C_\mu (X)\equiv \vec \mu (C(X))$ will be a rational cone in $\mathbb{R}^3$, called the moment map cone. Many geometrical properties of $C(X)$ as well as $X$ can be read off directly from $C_{\mu}(X)$.\\

The moment map cone can be specified by giving its inwards pointing normals $\{ \vec v_1,\ldots, \vec v_{\tt n} \}$, which we assume to be primitive (i.e. $\gcd (\vec v_i)=1 \ \forall i$). One can reconstruct $X$ from $C_{\mu}(X)$, as shown by Lerman \cite{2001math......7201L}. For $X$ to be smooth $C_{\mu}(X)$ must be \emph{good}.
The condition can be stated as follows \cite{Qiu:2014oqa}:
For a cone in $m$ dimensions, at every codimension $0<k<m$ face of the cone, the inwards facing normals of the (hyper) planes intersecting along it, $\{ \vec v_{i_1},\ldots,\vec v_{i_{k}} \}$, can be completed into an $SL(m,\mathbb{Z})$ matrix. In particular, for 5D manifolds $X$, one needs to check the goodness at the intersections of two codimension 1 faces.
That $X$ being SE (or $C(X)$ being CY) also translates to the so called \emph{1-Gorenstein condition} on the moment map cone. One way of stating this condition is that if there exists an integer vector $\vec \xi$ such that $\vec \xi \cdot \vec v_i = 1 \ \forall i$, then $C(X)$ is Calabi-Yau.
\\

A useful way of thinking about $C(X)$ is as a $T^3$ fibration over the interior of the moment map cone, i.e.  $T^3 \rightarrow X\to  C_\mu(X)^\circ$.  As we approach the 2d (resp. 1d, 0d) faces of $C_\mu (X)$, one (resp. two, three) of the torus fibers degenerate. The weight of the torus degenerating at a face is given by the norm(s) of the face.

\begin{figure}[h]
\begin{center}
\begin{tikzpicture}[scale=.8]
\draw [-,blue] (-1,-1) -- node[below] {\small$1$} (1,-1) -- node[right] {\small$2$} (1.5,.5) -- node[right] {\small$3$} (-.5,2) -- node[left] {\small$4$} (-1.5,1) -- node[left] {\small$5$} (-1,-1);

\draw (1,-1.6) ellipse (.3 and .6);
\draw (-1,-1.4) ellipse (.2 and .4);
\draw (1.5,0.8) ellipse (.18 and .3);
\draw (-0.5,2.3) ellipse (.1 and .3);
\draw (-2,1) ellipse (.5 and .2);

\draw [->,blue] (0,-1) -- (0,-.5) node[left] {\small$\vec v_1$};
\draw [->,blue] (1.25,-0.25) -- (0.95,-.15) node[left] {\small$\vec v_2$};
\draw [->,blue] (0.5,1.25) -- (0.2,.85) node[below] {\small$\vec v_3$};
\draw [->,blue] (-1,1.5) -- (-0.6,1.1) node[above] {\small$\vec v_4$};
\draw [->,blue] (-1.25,0) -- (-0.85,0.1) node[above] {\small$\vec v_5$};

\end{tikzpicture}

\caption{The polygon base of a polytope cone. Over the interior of the polygon there is a $T^3$ fiber, but over the faces the $T^3$ degenerates into $T^2$, which further degenerate over the vertices to $S^1$, drawn as the circles in the figure. These circles are the only generic closed Reeb orbits.}\label{fig:base}
\end{center}
\end{figure}
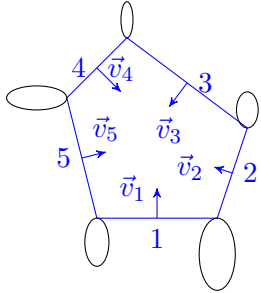

The Reeb vector $\reeb$ is a linear combination of the three torus actions, so we can naturally represent it as a vector $\vec \reeb\in\BB{R}^3$, i.e. $\vec\reeb$ gives the weights of $\reeb$. Then we obtain the actual manifold $X$ by restricting $C_\mu(X)$ to the plane $\vec\reeb\cdot\vec y = \frac 1 2$. We call this the base of the cone,
\[
	B_\mu (X) = \{ y \in C_\mu (X) | \vec y \cdot \vec \reeb = \frac 1 2 \}~,
\]
and if $\vec \reeb$ is within the dual cone of $C_\mu(X)$, then $B_\mu$ is a compact polygon, and $X$ is given by a $T^3$ fibration over the interior of the base, $T^3 \rightarrow B_\mu^\circ$.
An example, shown in figure \ref{fig:base} is the base of the so called $X^{p,q}$ manifold, where the different torus fibers degenerate as one moves towards the faces of the polygon, and at its vertices, only one $S^1$ remains.

For a generic Reeb (by generic, we mean that the ratios of the components of $\vec\reeb$ are irrational) its only closed orbits will be precisely located at the vertices of $B_{\mu}(X)$. As said already the neighborhood of each closed Reeb orbit can be identified with a solid torus $\mathbb{C}^2 \times S^1$, which is \emph{twisted}, i.e. we can present it as $\mathbb{C}^2 \times [0,2\pi \beta ] / \sim$, with the identification
\[
	(z_1, z_2 , 0) \sim ( e^{2\pi i\beta \epsilon} z_1, e^{2\pi i\beta \epsilon'} z_2 , 2\pi\beta )~,
\]
where $z_1,z_2$ are coordinates on $\mathbb{C}^2$. The twisting parameters of the solid tori are $\beta \epsilon$ and $\beta \epsilon'$, $\beta$ being the radius of the closed orbit. The parameters $\beta,\epsilon$ and $\epsilon'$ all depend on the Reeb vector, and appear in the factorised form of the partition function.
\\

We proceed to explain how to extract certain $SL(3, \BB{Z})$ matrices out of the moment map cone, and in particular, how to extract $\gb,\ep,\ep'$ for each closed Reeb orbit. Consider one of the vertices of $B_\mu(X)$ (corresponding to a 1d face of $C_\mu(X)$). At the vertex $i$, the two 2D faces with inward normals $\vec v_i,\vec v_{i+1}$ intersect, and we can order the two normals of the two 2d faces intersecting there such that $\det[\vec x, \vec v_i,\vec v_{i+1}] > 0$, where $\vec x$ is the generator of the 1d face. With this ordering and using the goodness of the cone, we can find a vector $\vec n_i$ such that $\det [ \vec n_i,\vec v_i,\vec v_{i+1} ] = 1$, meaning that these three vectors form an $SL(3,\BB{Z})$ basis. The two vectors $\vec v_i$ and $\vec v_{i+1}$, being the normal to the two faces, represent the torii has degenerated whereas $n_i$ represents the weight of the remaining non-degenerate circle. Expressing the Reeb vector in terms of this basis gives us
\be
	\vec \reeb = \frac{1}{\beta_i} \vec n_i + \epsilon_i \vec v_i + \epsilon_i' \vec v_{i+1}~,
\ee
where physically $\beta_i$ represents the radius of the $S^1$ and $\epsilon_i,\epsilon_i'$ are the equivariant rotation parameters. This equation directly gives us the relation
\be
	\begin{pmatrix} \beta_i^{-1} \\ \epsilon_i \\ \epsilon_i' \end{pmatrix} = [ \vec n_i, \vec v_i, \vec v_{i+1} ]^{-1} \vec \reeb \equiv \tilde K_i \vec \reeb~,
\ee
where we defined for vertex $i$ of the base $B_{\mu}(X)$, an $SL(3,\mathbb{Z})$ matrix,
\bea \label{eq:Kidef}
	\tilde K_i = [ \vec n_i,\vec v_i,\vec v_{i+1} ]^{-1}~.\label{def_K}
\eea
Written out explicitly
\bea
	\beta^{-1}_i &=& (\tilde K_i\vec\reeb)_1=\det [\vec  v_i,\vec  v_{i+1},\vec \reeb ],\nn\\
\epsilon_i &=& (\tilde K_i\vec\reeb)_2=\det [ \vec n_i, \vec \reeb, \vec v_{i+1} ]~,\\
\epsilon'_i &=& (\tilde K_i\vec\reeb)_3=\det [ \vec v_i, \vec \reeb,\vec n_i ] ~.\nn \label{identif_equiv_par}\eea

\subsection{Factorisation of the perturbative partition function}

In this section we write down the perturbative partition function for a simply connected 5D toric SE manifold $X$.
These manifolds have a canonically associated Reeb vector $\vec \reeb$, and we also denote by $\vec \xi$ the 3-vector such that $\vec\xi\cdot \vec v_i=1$ for all inwards pointing normals $v_i$ of the moment map cone $C_\mu(X)$.
The partition function for the Abelian theory coupled to an adjoint hypermultiplet can be calculated using the procedure outlined in \cite{Qiu:2014oqa} based on supersymmetric localization, and the result can be expressed in a concise way as
\be \label{eq:Zpert}
	Z^{\mathrm{pert}}_X = \frac{S_3^{C_\mu(X)'} ( 0 | \vec \reeb )}{S_3^{C_\mu (X)} (iM + \vec \xi \cdot \vec \reeb/2| \vec \reeb ) }~,
\ee
where $S_3^C$ is the generalized triple sine function associated to a good cone $C$, and the prime means that we remove the zero mode. The generalized triple sine associated to $C$ can be written as a product over all integer points inside $C$:
\bea
	S_3^C ( z | \vec \omega ) = \prod_{n\in C\cap\mathbb{Z}^3 } ( z + n\cdot\vec \omega ) \prod_{n\in C^\circ\cap\mathbb{Z}^3 } ( -z + n\cdot\vec \omega )~,\label{gen_sine}
\eea
where $C^\circ$ is the interior of $C$ and the product is understood to be $\zeta$-regulated, see \cite{Qiu:2014oqa,Tizzano:2014roa} for details.
Notice that for $S^5$, the moment map cone is $\mathbb{R}^3_{\geq 0}$ and $S_3^C$ becomes the usual triple sine function as we expect from \cite{Lockhart:2012vp}. \\

The generalised triple sine function comes from the fact that in the localization computation we compute a super determinant of the operator $-iL_{\sreeb}+\gs$ over the $\bar\partial_H$-complex. Here $L_{\sreeb}$ stands for Lie derivative along the Reeb vector. This operator shows up as the square of supersymmetry of cohomology complex  for the vector multiplet. For the hypermultiplet the operator $-iL_{\sreeb}+\gs$ is shifted by the mass $iM$ and also $\vec\xi\cdotp\vec\reeb/2$ arising from the difference of spin of the hyper complex compared with the vector complex.
Notice that in the Abelian theory one sets the Coulomb branch parameter $\gs$ to zero, and $-iL_{\sreeb}+\gs$ has a zero mode which must be excluded by hand. This explains the prime on $S_3^{C_\mu(X)'} ( 0 | \vec \reeb )$ in \eqref{eq:Zpert}.
In what follows, instead of excluding the zero mode we set $\gs=\gd$ and use $\gd$ as a regulator.
 Non-zero $\gd$ allows us to write our results in a simple factorised form, without the complications introduced by excluding a zero mode. Presumably we can think of non-zero $\gd$ as turning on a some sort of background field   in order to regularize the theory.
 \\

The partition function \eqref{eq:Zpert} is shown in \cite{Qiu:2014oqa} to factorise:
\be \begin{split}
\eqref{eq:Zpert} = &\exp\Big(-\frac{\pi i}{6}B_{3,3}^{C_{\mu}(X)}(\gd|\vec \reeb)+\frac{\pi i}{6}B_{3,3}^{C_\mu(X)}(iM+\vec \xi \cdot \vec \reeb /2 | \vec \reeb )\Big)\\ \times
&\prod_i^{\tt n} \frac{( e^{2\pi i\gb_i\gd} | e^{2\pi i \beta_i \epsilon_i } , e^{2\pi i \beta_i \epsilon_i' } )_\infty }{( e^{2\pi i\gb_i (iM + \vec \xi \cdot \vec \reeb/2) } | e^{2\pi i \beta_i \epsilon_i } , e^{2\pi i \beta_i \epsilon_i' } )_\infty }~ ,\label{eq:Zpert_fact}
\end{split} \ee
where we used the definition \eqref{identif_equiv_par} for the various parameters contained in the expression.
Here $B_{3,3}^{C_\mu}$ are the so called generalized Bernoulli polynomials, which are defined in appendix \ref{app:special}. They are third order polynomials of their first argument, and depend on the geometry of the cone (and thus of the 5d manifold $X$).
We introduce the notation
\[
 (z|\vec\reeb)_{\infty}^g=\big(\frac{z}{(g\vec\reeb)_1}\big|\frac{(g\vec\reeb)_2}{(g\vec\reeb)_1},\frac{(g\vec\reeb)_3}{(g\vec\reeb)_1}\big)_{\infty}~,~~~g\in SL(3,\BB{Z})~,
 \]
where we use the abbreviation $(e^{2i\pi z}|e^{2i\pi a},e^{2i\pi b})_{\infty}\to (z|a,b)_{\infty}$ to keep it more readable when the arguments get a bit too lengthy. Here, the group element $g\in SL(3,\mathbb{Z})$ acts on $(z|\vec\reeb)$ as a modular transformation.

With this notation and the matrices \eqref{def_K} defined from the cone, we can write the perturbative partition function (\ref{eq:Zpert_fact}) in a more suggestive form:
\bea Z_X^{\rm pert}=\big({\rm Bernoulli~factor}\big) \prod_i^{\tt n} \frac{(\gd|\vec\reeb)^{\tilde K_i}_\infty }{(iM + \vec \xi \cdot \vec \reeb/2|\vec\reeb)^{\tilde K_i}_\infty }~.\label{pert_fact_nice}\eea
This emphasizes that the various Nekrasov blocks are multiplied together with an appropriate $SL(3,\BB{Z})$ transformation, read off from the geometry of the cone.\\

As a simple example, consider $S^5$. The cone is the first octant in $\BB{R}^3$, and the three normals are $[1,0,0],[0,1,0]$ and $[0,0,1]$. Thus at vertex 1, the matrix is
\bea \tilde K_1=\left(
                  \begin{array}{ccc}
                    0 & 0 & 1 \\
                    1 & 0 & 0 \\
                    0 & 1 & 0 \\
                  \end{array}\right)\label{used_later_I} \ , \eea
and the corresponding block reads
\bea
(z|\vec\reeb)^{\tilde K_i}_{\infty}=(\frac{z}{\go_3}|\frac{\go_1}{\go_3},\frac{\go_2}{\go_3})_{\infty}~.\nn
\eea
The other two blocks are permutations of this, and we find the usual factorisation property of the regular multiple sine function \cite{MR2101221}.

\subsection{Constructing the full partition function}
Starting from the rewriting that we made in equation \eqref{eq:ZfullC2S1} we know that the full partition function on flat space is, up to some ambiguities, given by $G_2$ functions:
\bea Z^{\rm full}_{\BB{C}^2\times_{\ep} S^1}\sim\frac{G_2(\gb\gd|\ep,\ep',-\gb)}{G_2(\gb(iM+\vec\xi\cdotp\vec\reeb/2)|\ep,\ep',-\gb)}~ .
\nn\eea
Inspired by the factorisation result in the perturbative case, to get the full partition function we would like to multiply together copies of $Z^{\rm full}_{\BB{C}^2\times_{\ep} S^1}$, one coming from each distinct closed Reeb orbits in the geometry. For this we embed the matrices \eqref{def_K} into the $SL(4,\mathbb{Z})$ through
\[
	K_i = \begin{pmatrix}
		\tilde K_i & 0 \\
		0 & 1
	\end{pmatrix}~.
\]
Let us also denote by $S$ the 'S-duality' element in $SL(4,\mathbb{Z})$, i.e.
\be \label{eq:Smatrix}
	S = \begin{pmatrix}
		0 & 0 & 0 & -1 \\
		0 & 1 & 0 & 0 \\
		0 & 0 & 1 & 0 \\
		1 & 0 & 0 & 0
	\end{pmatrix}~.	
\ee
The element $S$ together with $SL(3,\mathbb{Z})$ generate all of $SL(4,\mathbb{Z})$.
We let $SL(4,\mathbb{Z})$ act as a fractional linear transformation on the parameters $(z | \omega_1, \omega_2,\omega_3)$ in the following way.
Introduce $\tilde \omega = (\omega_1, \omega_2,\omega_3,1)$ as an embedding of $\vec \go $ into $\mathbb{P}^3$, and then define the group action
\be  \label{eq:groupaction}
	g \cdot ( z | \vec \omega ) = \left ( \frac{z}{ (g\tilde \omega )_4 } | \frac{ (g\tilde\omega)_1 } { (g\tilde \omega )_4 } , \frac{ (g\tilde \omega )_2 }{ (g\tilde \omega )_4 }, \frac{(g\tilde\omega)_3}{ (g\tilde \omega )_4 } \right )~,~~~g\in SL(4,\BB{Z})~,
\ee
where $g\tilde \omega$ denotes ordinary matrix multiplication, and $(~)_i$ denotes the $i^{th}$ component.
We also let
\bea (g^*G_2)(z|\vec\reeb)=G_2\big(g\cdot(z|\vec\reeb)\big)\nn\eea
be the pull back of $G_2$ by the map induced by the $g$-action.
Now we propose the following full partition function
\bea
	Z^{\mathrm{full}}_X =\big(\textrm{Bernoulli factor}\big) \prod_{i=1}^{\tt n} (SK_i)^* \left ( \frac{  G_2 ( \gd | \vec \reeb ) } { G_2 ( iM+\vec\xi\cdot\vec\reeb/2|\vec \reeb ) } \right )~,\label{attempt_I}
\eea
where the missing Bernoulli polynomials of the first factor will be determined shortly. The physical idea behind such a factorisation is that we can interpret each block as contributions from widely separated pointed instanton particles propagating along the closed Reeb orbits. Unfortunately, at the moment, we are not able to derive this statement from first principles.
\\

As a guiding example, we consider $S^5$ again. At the vertex 1, the matrix $K_1$ is extended from \eqref{used_later_I}, and so
\bea
 (SK_1)(\gd|\vec\reeb)=\big(\frac{\gd}{\go_3}|-\frac{1}{\go_3},\frac{\go_1}{\go_3},\frac{\go_2}{\go_3}\big)~,\nn
 \eea
and the remaining factors of \eqref{attempt_I} are obtained by cyclic permutations.

Now we invoke an important modularity property the double elliptic gamma functions $G_2$ enjoys:  
\be\label{eq:modul}
\begin{split}
	G_2 ( z | \omega_1,\omega_2,\omega_3 ) &= e^{\frac{\pi i }{12} B_{4,4} ( z | \omega_1,\omega_2,\omega_3,-1)}\;\times  \\
	&G_2 ( \frac{z}{\omega_1} | \frac{\omega_2}{\omega_1},\frac{\omega_3}{\omega_1},-\frac {1 } {\omega_1} )
	G_2 ( \frac{z}{\omega_2} | \frac{\omega_1}{\omega_2},\frac{\omega_3}{\omega_2},-\frac {1 } {\omega_2} )
	G_2 ( \frac{z}{\omega_3} | \frac{\omega_1}{\omega_3},\frac{\omega_2}{\omega_3},-\frac {1 } {\omega_3} )~.
	\end{split}
\ee
Guided by this relation , which we view as the factorisation result for the standard cone, we propose that the missing Bernoulli factor in \eqref{attempt_I} in the case of $S^5$ is
\[ e^{\frac{i\pi}{12} ( B_{4,4} ( \delta | \omega_1,\omega_2,\omega_3,-1) - B_{4,4} (  iM+\vec\xi\cdot\vec\reeb/2 | \omega_1,\omega_2,\omega_3,-1) )}~, \]
where $B_{4,4}$ is a Bernoulli polynomial.
Consequently $Z_{S^5}^{\rm full}$ is also written in terms of $G_2$:
\bea
	 Z^{\mathrm{full}}_{S^5}=\frac{G_2(\gd|\go_1,\go_2,\go_3)}{G_2(iM+\vec\xi\cdotp\vec\reeb/2|\go_1,\go_2,\go_3)}~,\nn
	 \eea
where $\vec\xi\cdotp\vec\reeb=(\go_1+\go_2+\go_3)/2$. This is of course a previously known result of Lockhart and Vafa \cite{Lockhart:2012vp}, which we now will go on to generalize.

%
%

\subsection{Properties of the full partition function}\label{sec_Potfpf}
The information contained in equation \eqref{attempt_I} can be repackaged into a new special function which generalizes the double elliptic gamma function to take into account the geometry of the cone where we want to study our theory. This function was called generalized double elliptic gamma function in \cite{Tizzano:2014roa} and we can think about it in the same way as we think about the generalized the triple sine \eqref{gen_sine}, i.e. as a product over the cone:
\bea
	G_2^C ( z | \vec \omega ) = \prod_{\vec n \in C\cap \mathbb{Z}^3 } ( 1 - e^{2\pi i ( z + \vec n \cdot \vec \omega ) } ) \prod_{\vec n \in C^\circ \cap \mathbb{Z}^3 } ( 1 - e^{2\pi i ( -z + \vec n \cdot \vec \omega ) } )~ .\label{def_G_2}
\eea
 Due to the need to regulate the infinite product, we require $\im \go$ to lie within the dual cone $C^{\vee}$.
Taking the cone $C$ to be the first octant in $\BB{R}^3$, corresponding to $S^5$, one recovers the standard $G_2$.
This new $G_2^{C}$ has the factorisation property
\be
 	G_2^C ( z | \vec\reeb ) = e^{\frac{\pi i}{12} B_{4,4}^{\hat C} ( z | \vec\sreeb , -1) } \prod_{i=1}^{\tt n} (SK_i)^* G_2 ( z | \vec\reeb )~,\label{used_later_II}
\ee
where $B_{4,4}^{\hat C}$ is the Bernoulli polynomial associated with a \emph{4-dimensional} cone $\hat C$, whose normals are $\{(\vec v_i,0),(\vec 0,1)|$ $i=1,\cdots,\tt n\}$, where $\vec v_i$ are the normals of $C$, i.e. $\hat C = C \times \mathbb{R}_{\geq 0}$. The  proof  of (\ref{used_later_II})  is worked out in \cite{Tizzano:2014roa} and summarized in appendix \ref{app:special}.
It is quite curious that the parameters inside the various blocks, that we need to multiply together to get the full answer, enjoy the use of $SL(4,\BB{Z})$ matrices instead of the $SL(3,\BB{Z})$ ones. It would be nice to understand if there is any physical meaning about this observed modularity property.\\


For a simply connected toric SE manifold $X$, with moment map cone $C$, the full partition function for the $\mathcal{N}=2$ theory can be written in the very concise form
\be
	Z^{\rm full}_X = \frac{G_2^{C} ( \gd | \vec\reeb ) } { G_2^{C} (iM+\vec\xi\cdot\vec\reeb/2 | \vec \reeb ) }~.\label{used_later_III}
\ee
Given that the full abelian partition function on the five-sphere is related to the superconformal index of the theory living on the worldvolume of a single $M5$ brane, it would be nice to understand if also the Sasaki-Einsten case might have an index interpretation.

 Let us comment on the possible explanation for the appearance of $B_{4,4}$ factor. In the perturbative sector one has $B_{3,3}$ that leads to a $\gs^3$ term in the Coulomb branch. This matches the 1-loop generation of the Chern-Simons term $CS_5$ (see e.g. \cite{Seiberg:1996bd}). More precisely, one finds the $\gs^3$ term amongst the supersymmetric completion of $CS_5$ on curved space, with coefficients given by that of $B_{3,3}$. If we naively replace $\gs^3$ with $\sum_n(\gs+n)^3$, i.e. we sum over Kaluza-Klein modes, then $\sum_n(\gs+n)^3\sim \gs^4$ matching the coefficient of $B_{4,4}^{\hat C}$.

\section{Gluing rules}\label{sec:conemodularity}
We have shown that the full partition function is given by some $G_2$'s associated with certain cones as in (\ref{used_later_III}), and it is therefore just a weighted product over all lattice points inside the said cone. In this section we present another way to construct the full partition function in terms of elementary blocks glued together according to a set of rules. We state upfront that combinatorially these rules are \emph{tautological}: in that a weighted product taken in a cone can equally be done by first subdividing the cone  into smaller ones and taking the product in each sub-cone. Overcounting can arise since lattice points on shared boundaries are counted multiple times, and our set of rules is just a way to keep track of such overcounting.

\begin{figure}[h]
\begin{center}
\begin{tikzpicture}[scale=.8]
\draw [-,blue] (-0.1,0) -- (-1.1,0) -- (-2.1,-1) -- (-1.1,-2) -- (-.1,0);
\node at (0,0) {\small$\bullet_1$};
\node at (0.4,.4) {\small$(0,0)$};
\node at (-1,0) {\small$\bullet_2$};
\node at (-1,.4) {\small$(-1,0)$};
\node at (-2,-1) {\small$\bullet_3$};
\node at (-2,-1.4) {\small$(-2,-1)$};
\node at (-1,-2) {\small$\bullet_4$};
\node at (-1,-2.4) {\small$(-1,-2)$};
%
%
\end{tikzpicture}
\caption{The toric fan of $C(Y^{2,1})$, not to be confused with figure \ref{fig_subdiv_Y21} or \ref{fig_triangulation}.}\label{fig:toric_Y21}
\end{center}
\end{figure}
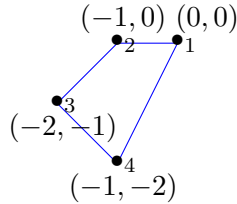

As the construction pivots on subdivision of cones, we would like to start by clarifying a possible confusion in this regard.
As is typical in the literature, for instance see \cite{Aganagic:2003db}, toric CY 3-folds are represented by the toric fans (see \cite{Fulton} for a fuller exposition), which are made of cones fulfilling certain conditions. The generators of the cones, which are primitive vectors of $\BB{Z}^3$ can be chosen so that their first component is 1. Thus one often draws only the 2nd and 3rd components of generators. For example for $C(Y^{p,q})$, the generators are
\bea v_1=[1,0,0],~~v_2=[1,-1,0],~~v_3=[1,-2,-p+q],~~v_4=[1,-1,-p]\label{normal_Ypq}\eea
see figure \ref{fig:toric_Y21}.
In this presentation, resolution of singularity amounts to subdividing the cones and one often uses the trivalent vertex to represent the toric diagram in e.g. the topological vertex formulation.  We want to stress
that the subdivision of this paragraph is \emph{not} to be confused with the subdivision which we discuss later. To be more precise, for this paper we always work with the moment map cone, whose \emph{faces have normals} $\vec v_i$. The subdivision of the moment map cone corresponds to gluing toric contact manifolds together along their common boundary which are represented by the added faces.

\subsection{Subdivision of the cone and gluing construction of toric 5d manifolds} \label{sec:gluingcons}
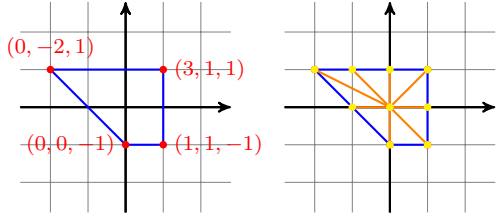
\begin{figure}
\begin{center}
\begin{tikzpicture}[fill opacity=1, draw opacity=1]
	\draw[step=0.5cm,gray,very thin] (-1.4,-1.4) grid (1.4,1.4);
	\draw[thick, ->] (-1.4, 0) -- (1.4,0);
	\draw[thick, ->] (0, -1.4) -- (0,1.4);
	\draw[thick,blue] (-1,0.5) -- (0.5, 0.5) -- (0.5, -0.5) -- (0, -0.5) -- cycle;
		\fill [red](-1, 0.5) circle (0.05) node[above]{\scriptsize$(0,-2,1)$};
		\fill [red](0.5, 0.5) circle (0.05) node[right]{\scriptsize$(3,1,1)$};
		\fill [red](0.5,-0.5) circle (0.05) node[right]{\scriptsize$(1,1,-1)$};
		\fill [red](0, -0.5) circle (0.05) node[left]{\scriptsize$(0,0,-1)$};	
\end{tikzpicture}
\begin{tikzpicture}[fill opacity=1, draw opacity=1]
	\draw[step=0.5cm,gray,very thin] (-1.4,-1.4) grid (1.4,1.4);
	\draw[thick, ->] (-1.4, 0) -- (1.4,0);
	\draw[thick, ->] (0, -1.4) -- (0,1.4);
	\draw[thick,blue] (-1,0.5) -- (0.5, 0.5) -- (0.5, -0.5) -- (0, -0.5) -- cycle;
		\fill [red](-1, 0.5) circle (0.05);
		\fill [red](0.5, 0.5) circle (0.05);
		\fill [red](0.5,-0.5) circle (0.05);
		\fill [red](0, -0.5) circle (0.05);
	\draw[thick, orange] (0,0) -- (0.5,-0.5);
	\draw[thick, orange] (0,0) -- (0,-0.5);
	\draw[thick, orange] (0,0) -- (0.5,0.5);
	\draw[thick, orange] (0,0) -- (0.0,0.5);
	\draw[thick, orange] (0,0) -- (0.5,0.0);
	\draw[thick, orange] (0,0) -- (-0.5,0.5);
	\draw[thick, orange] (0,0) -- (-1,0.5);
	\draw[thick, orange] (0,0) -- (-0.5,0);
		\fill [yellow](0.5,-0.5) circle (0.05);
		\fill [yellow](0., -0.5) circle (0.05);
		\fill [yellow](0.5,0.5) circle (0.05);
		\fill [yellow](0, 0.5) circle (0.05);
		\fill [yellow](0.5, 0) circle (0.05);
		\fill [yellow](-0.5, 0.5) circle (0.05);
		\fill [yellow](-1, 0.5) circle (0.05);		
		\fill [yellow](-0.5, 0) circle (0.05);		
		\fill [yellow](-0., 0.) circle (0.05);
\end{tikzpicture}
\caption{The left panel is the base $B$ of the moment map cone of $Y^{2,1}$, the four normals are given in \eqref{normal_Ypq}. A point $(m,n)$ in the grid represents a vector $(1,0,0)+m(1,1,0)+n(1,0,1)$. The right panel gives a subdivision of the cone.}
\label{fig_subdiv_Y21}
\end{center}
\end{figure}

It is always possible to subdivide a good 3D cone $C$ such that each smaller cone has three faces and the three normals form $SL(3,\BB{Z})$ basis \cite{Fulton}, furthermore, two such cones intersect at a collection of lower dimensional faces common to both (put more simply, the subdivision leads to a simplicial subdivision of the base of the cone).
We call these cones \emph{good simplicial cones}. Since we constantly draw the cone $C$ by drawing its base $B$, the subdivision looks like figure \ref{fig_subdiv_Y21}.
%
%
%
%
Recall that the $G^C_2$ function is a weighted product over lattice points of the cone $C$. If $C$ is a good simplicial cone, then it can be turned into the first octant of $\mathbb{R}^3$ by an $SL(3,\BB{Z})$ transformation and so the product gives the standard $G_2$ with some transformed parameters.
For a general $C$, we subdivide it into good simplicial cones $\{C_i\}$ and the product \eqref{def_G_2} within $C_i$ gives a $G_2^{C_i}$ for each $i$.
The overcounting mentioned above arises when, say, $C_i,\,C_j$ share a 2D face, and so the lattice points on this face are counted twice, and must be corrected. Similarly the overcounting will also arise when a number of cones share a 1D face. The following set of rules gives us a systematic way of removing the overcounting.

\begin{figure}[h]
\begin{center}
%
%
%
\begin{tikzpicture}[scale=1.5,fill opacity=1, draw opacity=1]
	\draw[thick,blue] (-1,0.5) -- (0.5, 0.5) -- (0.5, -0.5) -- (0, -0.5) -- cycle;
	\draw[thick, orange] (0,0) -- (0.5,-0.5);
	\draw[thick, orange] (0,0) -- (0,-0.5);
	\draw[thick, orange] (0,0) -- (0.5,0.5);
	\draw[thick, orange] (0,0) -- (0.0,0.5);
	\draw[thick, orange] (0,0) -- (0.5,0.0);
	\draw[thick, orange] (0,0) -- (-0.5,0.5);
	\draw[thick, orange] (0,0) -- (-1,0.5);
	\draw[thick, orange] (0,0) -- (-0.5,0);
	\draw (0.16,0.33) -- (-.16,.36) -- (-.5,.33) -- (-.5,.16) -- (-.16,-.16) -- (.16,-.33) -- (.33,-.16) -- (.33,.16) -- cycle;
    \draw (0.16,0.33) -- (.16,.66);	
    \draw (-.16,.36) -- (-.16,.66);	
    \draw (-.5,.33) -- (-1,.66);
    \draw (-.5,.16) -- (-1,.16);
    \draw (-.16,-.16) -- (-.48,-.48);
    \draw (.16,-.33) -- (.16,-.66);
    \draw (.33,-.16) -- (.66,-.16);
    \draw (.33,.16) -- (.66,.33);
\end{tikzpicture}
\end{center}
\caption{Triangulation of the cone of $Y^{2,1}$ and the dual 3-valent graph. We warn the reader that this is \emph{not} related to resolution of singularity, see the beginning of section \ref{sec:conemodularity}.}\label{fig_triangulation}
\end{figure}
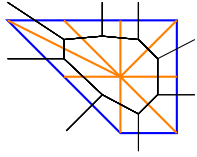

We draw only the base $B$ of the cone $C$, and the subdivision of $C$ leads to a simplicial subdivision of $B$, which we represent using tri-valent graphs, see fig.\ref{fig_triangulation}. In this figure each 2-simplex is dual to a tri-valent vertex, each 1-simplex is dual to an edge and an internal 0-simplex will be dual to a loop.
We borrow a terminology from the topological vertex \cite{Aganagic:2003db} and describe a set of rules for the gluing procedure
\begin{itemize}
\item To every trivalent vertex we associate a \emph{vertex} factor
 \be
 	V( M | \underline \go_v ) = \frac{ G_2( \gd | \underline \go_v ) } { G_2 ( iM + \reeb \cdot \xi/2 | \underline \go_v ) }~ .\label{vertex}
 \ee	
\item To each internal edge connecting two vertices, we associate a \emph{propagator} factor
 \bea
 	P( M | \underline \go_p ) = \frac{G_1 ( \gd | \underline \go_p ) }{ G_1 (  iM + \reeb \cdot \xi/2  | \underline \go_p ) }~ .\label{Edge}
 \eea
 \item Finally if the tri-valent graph has closed loops, we insert for each of them a \emph{loop} factor
\bea
	L(M | \go_l ) = \frac{ G_0(\gd| \go_l )} { G_0 (  iM + \reeb \cdot \xi/2  | \go_l ) }~.\label{Loop}
\eea
\end{itemize}
The parameters $\underline \go_v, \underline\go_p,\go_l$ are given as follows.
Each edge of a trivalent vertex has an associated normal vector (the normal of the face this edge 'pierces'), and there is an ordering such that the normals form an $SL(3,\mathbb{Z})$ matrix $A_v^{-1} = [ v_1,v_2,v_3 ]$.
The parameters associated to this vertex are then $\underline\go_v = A_v \vec\reeb$.
Equivalently, one can let $x_1,x_2,x_3$ be the three generators of the cone associated to the trivalent vertex, then one may write $\underline \go_v = (x_1\cdot \reeb , x_2\cdot\reeb, x_3\cdot\reeb)$\footnote{if $x_i$ is the primitive generator along the intersection of two faces with normal $v_i$ and $v_{i+1}$ then clearly $x_i=v_i\times v_{i+1}$, and one has $A_v=[v_1,v_2,v_3]^{-1}=[x_1,x_2,x_3]$.}. 

The propagators (internal edges) are dual to 2d cones, each such cone has two associated generators, say, $x_1, x_2$, then $\underline{\go}_p$ is given by $ \underline \go_p = (x_1 \cdot \vec\reeb , x_2 \cdot \vec\reeb)$.

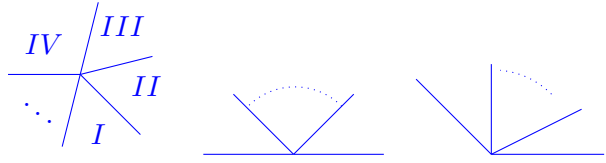
\begin{figure}[h]
\begin{center}
\begin{tikzpicture}[scale=0.8]
\draw [-,blue] (0,0)--(-1.2,0);
\draw [-,blue] (0,0) -- (-0.3,-1.2);
\draw [-,blue] (0,0) -- (1,-1);
\draw [-,blue] (0,0) -- (1.2,0.3);
\draw [-,blue] (0,0) -- (0.3,1.2);

\node[blue] at (-.7,-.5) {\small$\ddots$};
\node[blue] at (0.3,-1) {\small$I$};
\node[blue] at (1.1,-.2) {\small$II$};
\node[blue] at (.7,.8) {\small$III$};
\node[blue] at (-.6,0.5) {\small$IV$};
\end{tikzpicture}
~
\begin{tikzpicture}[scale=0.8]
\draw [-,blue] (0,0)--(1.5,0);
\draw [-,blue] (0,0)--(1,1);

\draw [dotted,blue] (0.8,0.8) arc (45:135:1.1);

\draw [-,blue] (0,0) -- (-1,1);
\draw [-,blue] (0,0) -- (-1.5,0);
\end{tikzpicture}
~
\begin{tikzpicture}[scale=1]
\draw [-,blue] (0,0)--(1.5,0);
\draw [-,blue] (0,0)--(1.2,.6);

\draw [dotted,blue] (0.8,0.8) arc (45:85:1.1);
\draw [-,blue] (0,0) -- (0,1.2);
\draw [-,blue] (0,0) -- (-1,1);
\end{tikzpicture}
\caption{A number of good simplicial cones conjoining along a 1D cone. The first picture, the 1D cone is in the interior of $C$; in the second, it is on a 2D face of $C$; and at a 1D face of $C$ in the third picture.}\label{fig_conjoin}
\end{center}
\end{figure}

Finally for the loop factor, which is dual to a shared internal 1d face with generator $x_l$, and the corresponding parameter is given by $\go_l = x_l \cdot \reeb$.
Putting together the $V,P,L$ factors, one removes all the overcounting and each lattice point in the cone contributes exactly once to the product in $G_2^C$.
Indeed, referring to the first picture of fig.\ref{fig_conjoin} where $n$ good simplicial cones conjoin at an internal 1D cone, then the shared faces are double counted and the contribution \eqref{Edge} cancels the double count, as all the shared faces are dual to internal edges. In this process, the 1D cone in the middle is counted $n$ times in \eqref{vertex}, also removed $n$ times by \eqref{Edge}, and finally \eqref{Loop} restores it, since internal 1D cone are dual to loops. By contrast, in the second picture of fig.\ref{fig_conjoin} where $n$ good simplicial cones conjoin along a 1D cone which is in the face of the original cone. Then the two outer faces are not overcounted, but we do not assign \eqref{Edge} to these two faces since they are not dual to internal edges. The shared 1D cone is counted $n$ times by \eqref{vertex} but removed $n-1$ times by \eqref{Edge} and hence counted correctly. Indeed we do not assign \eqref{Loop} since this 1D cone is not dual to a loop. The third picture of fig.\ref{fig_conjoin} is treated in the same way.

As said, these rules are quite tautological in itself, but we can also view the subdivision as a cutting, by means of an appropriate surgery, of the 5d manifold into pieces of topology $S^5$. Each $S^5$ contributes \ref{vertex}, since as we recall the moment map cone of $C(S^5)=\BB{C}^3$ is a good simplicial cone.
At the level of the path integral, the cutting is achieved by adding defects along the shared boundaries, with its own degrees of freedom that contribute factors of \eqref{Edge} and \eqref{Loop}. We will comment briefly about this possibility in section \ref{sec:gluingdefects}.


\subsection{Partition function for some simple $Y^{p,q}$}
In section \ref{sec_Potfpf} we constructed the full partition function by essentially reversing the factorisation of $G_2^C$ into Nekrasov partition functions associated to $\BB{C}^2\times_{\ep}S^1$. In section \ref{sec:gluingcons} we give a reconstruction of $G_2^C$ from some gluing rules. In this section we carry out the calculation on some simple $Y^{p,q}$'s in detail, including the factorisation, as an atonement for not giving a self-contained rigorous proof about the factorisation property of $G_2^C$.

We first sketch a proof of the factorisation of $G_2^C$. From the rules of previous section, one associates a $G_2^{C_i}$ factor to $C_i$. With $C_i$ being good simplicial, one can factorise $G_2^{C_i}$ according to \eqref{eq:G2modularity} into 3 standard $G_2$'s associated with the three 1D faces of the cone. We assemble these numerous $G_2$'s by grouping them according to the 1D cone they associate with. In this process massive cancellations will occur. In fact in the situation that a number of cones share a 1D cone we can list the following possibilities

\begin{itemize}
\item[1.] The shared 1D cone is in the interior of the original cone $C$.
\item[2.] The shared 1D cone is contained in a 2D face of $C$.
\item[3.] The shared 1D cone is also a 1D face of $C$.
\end{itemize}
In each of these situations, one has a product of $G_2$'s associated with the 1D cone. In the first two situations the net contribution cancels that of \eqref{Edge}, \eqref{Loop}. While for the third situation, the product gives the factors of \eqref{used_later_II}, one for each 1D cone of the original cone. This is essentially the proof of the factorisation \eqref{used_later_II}, and is the strategy used in \cite{Tizzano:2014roa} applied to cones associated to simply connected toric SE manifolds. In the proof there, the 1-Gorenstein condition is used, but the proof can also be extended to the case of all good cones, with an ingenious use of the integral representation of $G_m^C$. This will be published separately \cite{Winding:2016}.

\subsubsection{$Y^{1,0}$}
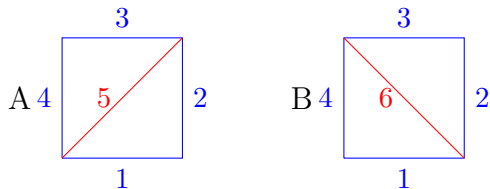
\begin{figure}[h]
\begin{center}
\begin{tikzpicture}[scale=.8]
\draw [-,blue] (-1,-1) -- node[below] {\small$1$} (1,-1) -- node[right] {\small$2$} (1,1) -- node[above] {\small$3$} (-1,1) -- node[left] {\small$4$} (-1,-1) ;
\draw[-,red](-1,-1) -- node[left]{\small $5$} (1,1) ;
\node (A) at  (-1.7,0.) {A} ;
\end{tikzpicture} \ \ \ \ \
\begin{tikzpicture}[scale=.8]
\draw [-,blue] (-1,-1) -- node[below] {\small$1$} (1,-1) -- node[right] {\small$2$} (1,1) -- node[above] {\small$3$} (-1,1) -- node[left] {\small$4$} (-1,-1) ;
\draw[-,red](1,-1) -- node[left]{\small $6$} (-1,1) ;
\node (A) at  (-1.7,0.) {B} ;
\end{tikzpicture}

\caption{The two possible triangular subdivisions of the moment map cone of $Y^{1,0}$. }\label{fig:conifoldgluing}
\end{center}
\end{figure}

The moment map cone over $Y^{1,0}$ has the following inward pointing normals:
\bea
	\vec v_1 = [1,1,-1]~,~~~~ \vec v_2 = [1,1,0]~, ~~~~ \vec v_3 = [1,0,1]~, ~~~~ \vec v_4 = [1,0,0]~.\eea
This choice of normals are equivalent to \eqref{normal_Ypq} up to a cyclic permutation and an $SL(2,\BB{Z})$ matrix $[-2,-1;-1,-1]$ acting on the 2nd and 3rd entries of $v_i$. As usual one has $\vec\xi\cdotp\vec v_i=1$ with $\vec\xi = [1,0,0]$. Let the Reeb be given by $\vec\reeb=[\go_1,\go_2,\go_3]$.
The subdivision of the cone into two good simplicial ones is done by adding another plane, either as in case A or as in case B indicated in figure \ref{fig:conifoldgluing}.
\begin{figure}
\begin{center}
\begin{tikzpicture}[scale=.8]
\draw [-,black] (-1,1.5) -- node[below] {\small$v_4$} (0,1.5) -- node[right] {\small$v_3 $} (0,2.5);
\draw[-,black](0,1.5) -- node[above] {\small$v_5 $} (1.5,0);
\draw[-,black](1.5,-1) -- node[left] {\small$v_1$} (1.5,0) -- node[below] {\small$v_2$} (2.5,0);
\node (A) at  (-1.7,0.) {A} ;
\end{tikzpicture} \ \ \ \ \
\begin{tikzpicture}[scale=.8]
\draw [-,black] (-1,0) -- node[below] {\small$v_4$} (0,0) -- node[right] {\small$v_1 $} (0,-1);
\draw[-,black](0,0) -- node[above] {\small$v_6 $} (1.5,1.5) -- node[below] {\small$v_2$} (2.5,1.5) ;
\draw[-,black](1.5,1.5) -- node[left] {\small$v_3$} (1.5,2.5);
\node (A) at  (-1.7,0.) {B} ;
\end{tikzpicture} \ \ \ \ \
\caption{The dual diagram of figure \ref{fig:conifoldgluing} showing the two different triangulations. } \label{fig:conifoldpq}
\end{center}
\end{figure}
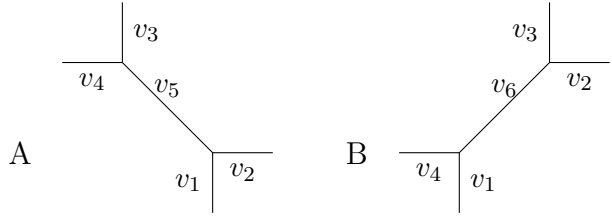

If we pick case A, the plane we add can be found to have the normal vector
\be
	\vec v_5 = [ 0,1,-1]~,
\ee
Let us consider the case $A$. The rule \eqref{vertex} gives two terms (we only take the numerator, as the treatment for the denominator is the same)
\[
	\begin{split}
	G_2 ( z | A_{125} \vec \reeb ) &= G_2 ( z | \go_1 - \go_2 -  \go_3, \go_2+\go_3,\go_1-\go_2 )~, \\
	G_2( z | A_{345}\vec \reeb) &= G_2 ( z |\go_1 - \go_2 - \go_3 ,\go_2+\go_3, \go_2 )~.
	\end{split}
\]
where $A_{125}= [\vec v_1,\vec v_2,-\vec v_5]^{-1}$ or $A_{345} = [\vec v_3,\vec v_4,\vec v_5]^{-1}$ respectively. The rule \eqref{Edge} gives
\bea G_1 ( z | [0,1,1]\cdot\vec \reeb , [1,-1,-1]\cdot \vec \reeb )=G_1 ( z | \go_2+\go_3, \go_1-\go_2-\go_3).\label{used_later_IV}\eea
In this situation there is no closed loop, and it is fairly obvious that the $G_1$ factor here cancels the double counting on the added face and altogether we recover $G_2^C(z|\vec\reeb)$.

Next we factorise the two $G_2$'s above. To emphasise the main message, we will suppress the Bernoulli factors. In the following we use $\sim$ to mean equal up to Bernoulli
\bea
	G_2 ( z | A_{125}\vec\reeb ) &\sim&
	G_2( \frac{z}{\go_1 - \go_2 } | \frac{-\go_3}{\go_1 - \go_2 },\frac{\go_2+\go_3}{\go_1 - \go_2 }, \frac{-1}{\go_1 - \go_2  } )G_2( \frac{z}{\go_2 + \go_3 } | \frac{\go_1}{\go_2 + \go_3 },\frac{\go_1-\go_2}{\go_2 + \go_3 }, \frac{-1}{\go_2 + \go_3  } ) \nn\\
	&&G_2 ( \frac{z}{\go_1-\go_2-\go_3} | \frac{\go_2+\go_3}{\go_1-\go_2-\go_3}, \frac{\go_1-\go_2}{\go_1-\go_2-\go_3} , \frac{-1}{\go_1-\go_2-\go_3} ),\nn\\
G_2 ( z | A_{345}\vec\reeb )&\sim&
	G_2 ( \frac{z}{\go_2} | \frac{\go_1-\go_3}{\go_2} , \frac{\go_3}{\go_2} , \frac{-1}{\go_2} )
	G_2 ( \frac{z}{\go_2+\go_3} | \frac{\go_1}{\go_2+\go_3},\frac{\go_2}{\go_2+\go_3}, \frac{-1}{\go_2 + \go_3})\nn\\
    &&G_2 ( \frac{z}{\go_1-\go_2-\go_3} | \frac{\go_2+\go_3}{\go_1-\go_2-\go_3}, \frac{\go_2}{\go_1-\go_2-\go_3} , \frac{-1}{\go_1-\go_2-\go_3} ),
	\label{used_later_V}
\eea
where we have also used the relation $G_2(z|a,b,c)=G_2(z|a+1,b,c)$ to simplify some of the terms. We will name the six $G_2$'s in \eqref{used_later_V} as $I,\cdots,VI$.

Applying the following property (see \cite{Tizzano:2014roa}) to the second $G_2$ above,
\bea
		G_2 ( z| a,a+b,c) G_2 ( z| a+b, b, c) = \frac{ G_2 ( z | a,b,c) } { G_1 ( z | a+b,c) } ~, 	\label{used_later_VII}
\eea
one obtains
\bea &&II
=G_2( \frac{z}{\go_2 + \go_3 } | \frac{\go_1-\go_2}{\go_2 + \go_3 },\frac{-\go_2}{\go_2 + \go_3 }, \frac{-1}{\go_2 + \go_3  } )^{-1}\nn\\
&&\hspace{2cm}G_2( \frac{z}{\go_2 + \go_3 } | \frac{\go_1}{\go_2 + \go_3 },\frac{-\go_2}{\go_2 + \go_3 }, \frac{-1}{\go_2 + \go_3  } )
G_1( \frac{z}{\go_2 + \go_3 } | \frac{\go_1-\go_2}{\go_2 + \go_3 }, \frac{-1}{\go_2 + \go_3  } )^{-1}.\nn\eea
This combines with the fifth $G_2$ of \eqref{used_later_V} to be (using $G_2(z|a,b,c)G_2(z|-a,b,c)=G_1(z|b,c)^{-1}$) one obtains
\bea II\times V=G_2( \frac{z}{\go_2 + \go_3 } | \frac{\go_1-\go_2}{\go_2 + \go_3 },\frac{\go_2}{\go_2 + \go_3 }, \frac{-1}{\go_2 + \go_3  } )
G_1( \frac{z}{\go_2 + \go_3 } | \frac{\go_1}{\go_2 + \go_3 }, \frac{-1}{\go_2 + \go_3  } ).\nn\eea
In the same manner the third and sixth $G_2$ in \eqref{used_later_V} combine into
\bea&& III\times VI=
G_1 ( \frac{z}{\go_1-\go_2-\go_3} | \frac{\go_1}{\go_1-\go_2-\go_3}, \frac{-1}{\go_1-\go_2-\go_3} )\nn\\
&&\hspace{3cm} G_2 ( \frac{z}{\go_1-\go_2-\go_3} | \frac{\go_3}{\go_1-\go_2-\go_3}, \frac{\go_2}{\go_1-\go_2-\go_3} , \frac{-1}{\go_1-\go_2-\go_3} )\nn\eea
The two $G_1$ terms cancel that of \eqref{used_later_IV} after factorising the latter. Finally we have proved the the special case of \eqref{used_later_II} up to Bernoulli's
\be	
	\begin{split}
	G_2^C ( z | \vec\reeb ) &\sim
	G_2( \frac{z}{\go_1 - \go_2 } | \frac{-\go_3}{\go_1 - \go_2 },\frac{\go_2+\go_3}{\go_1 - \go_2 }, \frac{-1}{\go_1 - \go_2  } ) \\
	&\times G_2 ( \frac{z}{\go_1-\go_2-\go_3} | \frac{\go_2}{\go_1-\go_2-\go_3}, \frac{\go_3}{\go_1-\go_2-\go_3} , \frac{-1}{\go_1-\go_2-\go_3} )
	G_2 ( \frac{z}{\go_2} | \frac{\go_3}{\go_2}, \frac{\go_1 - \go_3}{\go_2} , \frac{-1}{\go_2} ) \\
	&\times G_2 ( \frac{z}{\go_2+\go_3} | \frac{\go_1-\go_2}{\go_2+\go_3},\frac{\go_2}{\go_2+\go_3}, \frac{-1}{\go_2 + \go_3})	~.\nn
	\end{split}
\ee
We observe that the first $G_2$ factor is the Nekrasov partition function on $\BB{C}^2\times_{\ep}S^1$ associated with the corner (or 1d face) 12, where face 1 and 2 intersect, the second from the corner 23 and so on.

One can do the same computation by choosing the subdivision $B$, the result is of course the same.

\subsubsection{$Y^{2,1}$}
From the subdivision fig.\ref{fig_triangulation}, one gets a total of 8 simplicial cones and hence 8 $G_2$ functions. As performing the entire calculation will be rather lengthy, and so we highlight some feature that was not in the $Y^{1,0}$ case.
\begin{figure}[h]
\begin{center}
\begin{tikzpicture}[scale=2,fill opacity=1, draw opacity=1]
	\draw[thick,blue] (-1,0.5) -- (0.5, 0.5) -- (0.5, -0.5) -- (0, -0.5) -- cycle;
		\fill [red](-1, 0.5) circle (0.05);
		\fill [red](0.5, 0.5) circle (0.05);
		\fill [red](0.5,-0.5) circle (0.05);
		\fill [red](0, -0.5) circle (0.05);
		
	\fill[green](0,0) -- (0.5,0) -- (0.5, -0.5) -- cycle;
	\fill[purple](0,0) -- (0.5,0) -- (0.5,0.5) -- cycle;
	\draw[thick, orange] (0,0) -- (0.5,-0.5);
	\draw[thick, orange] (0,0) -- (0,-0.5);
	\draw[thick, orange] (0,0) -- (0.5,0.5);
	\draw[thick, orange] (0,0) -- (0.0,0.5);
	\draw[thick, orange] (0,0) -- (0.5,0.0);
	\draw[thick, orange] (0,0) -- (-0.5,0.5);
	\draw[thick, orange] (0,0) -- (-1,0.5);
	\draw[thick, orange] (0,0) -- (-0.5,0);
		\fill [yellow](0, 0.5) circle (0.05);
		\fill [yellow](0.5, 0) circle (0.05);
		\fill [yellow](-0.5, 0.5) circle (0.05);
		\fill [yellow](-0.5, 0) circle (0.05);		
		\fill [yellow](-0., 0.) circle (0.05);

    \node at (.15,.05) {\tiny{V}};
    \node at (.4,.05) {\tiny{IV}};
    \node at (.45,.4) {\tiny{VI}};
    \node at (.15,-.05) {\tiny{II}};
    \node at (.4,-.05) {\tiny{I}};
    \node at (.47,-.4) {\tiny{III}};
    \node at (.75,0) {\tiny{(2,1,0)}};
    \node at (.75,0.5) {\tiny{(3,1,1)}};
    \node at (.75,-0.5) {\tiny{(1,1,-1)}};
\end{tikzpicture}
\caption{The two simplicial cones whose contributions we consider. }
\label{fig:cones2}
\end{center}
\end{figure}
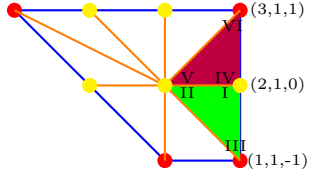

We consider the following two good simplical cones, the first one generated by the vectors $m_2 = [ 1,1,-1]$, $\tilde m = [ 2,1,0]$ and $m_5 = [ 1,0,0]$, corresponding to the green triangle in fig.\ref{fig:cones2}; and the second one generated by $m_5$, $\tilde m$ and $m_3 = [3,1,1]$, represented by the purple triangle in figure 2.
The first one gives us (up to Bernoulli polynomials, as usual)
\bea
    &&G_2(z|\go_1+\go_2-\go_3,2\go_1+\go_2,\go_1)\sim G_2 ( \frac{z}{2\go_1+\go_2} | \frac{\go_1}{2\go_1+\go_2}, \frac{-\go_1-\go_3}{2\go_1+\go_2}, \frac{-1}{2\go_1+\go_2})\nn\\
	&&\hspace{.5cm}
	G_2 ( \frac{z}{\go_1} | \frac{\go_2-\go_3}{\go_1}, \frac{\go_2}{\go_1}, \frac{-1}{\go_1} )
G_2 ( \frac{z}{\go_1+\go_2-\go_3} | \frac{2\go_1+\go_2}{\go_1+\go_2-\go_3}, \frac{\go_1}{\go_1+\go_2-\go_3}, \frac{-1}{\go_1+\go_2-\go_3} )\nn\eea
naming the three $G_2$'s on the rhs as $I,II,III$, that are associated to the corresponding corner marked in fig.\ref{fig:cones2}.
The second cone gives us similarly
\bea G_2(z|3\go_1+\go_2+\go_3,2\go_1+\go_2,\go_1)\sim G_2 ( \frac{z}{2\go_1+\go_2} | \frac{\go_1}{2\go_1+\go_2}, \frac{\go_1+\go_3}{2\go_1+\go_2}, \frac{-1}{2\go_1+\go_2} )~~~~~~~~~~~~~~\nn\\
G_2 ( \frac{z}{\go_1} | \frac{\go_2+\go_3}{\go_1}, \frac{\go_2}{\go_1}, \frac{-1}{\go_1} )G_2 ( \frac{z}{3\go_1+\go_2+\go_3} | \frac{\go_1}{3\go_1+\go_2+\go_3}, \frac{2\go_1+\go_2}{3\go_1+\go_2+\go_3}, \frac{-1}{3\go_1+\go_2+\go_3} )\nn\eea
Name the $G_2$'s here $IV,V,VI$ belonging to the corners marked in fig.\ref{fig:cones2}.

The $G_2$'s marked $I,\,IV$ fall into the case of the middle panel of fig.\ref{fig_conjoin}, which is a case that did not arise in the $Y^{1,0}$ case.
Indeed the propagator contribution \eqref{Edge} for the shared face between the purple and green cone gives
\bea G_1(z|\go_1,2\go_1+\go_2)\sim G_1(\frac{z}{\go_1}|\frac{\go_2}{\go_1})G_1(\frac{z}{2\go_1+\go_2}|\frac{\go_1}{2\go_1+\go_2}).\label{used_later_VI}\eea
The first $G_1$ will be grouped to the 'central hub' (with generator $m_5=(1,0,0)$). Grouping the second $G_1$ with $I$ and $IV$ gives
\bea G_1(\frac{z}{2\go_1+\go_2}|\frac{\go_1}{2\go_1+\go_2})\times I\times IV=G_1(\frac{z}{2\go_1+\go_2}|\frac{\go_1}{2\go_1+\go_2})G_1(\frac{z}{2\go_1+\go_2}|\frac{\go_1}{2\go_1+\go_2})^{-1}=1,\nn\eea
in accordance with the discussion below fig.\ref{fig_conjoin}.

Another new situation not in the $Y^{1,0}$ case is that we have a product of $G_2$'s the likes of $II$ and $V$, together with the $G_1$'s like the first one in \eqref{used_later_VI}. These $G_2$'s and $G_1$'s surround the central hub and so correspond to the left panel of fig.\ref{fig_conjoin}. And according to the discussion below that same figure, such contribution should be 1. To prove this, one can start by assuming that in the left panel of fig.\ref{fig_conjoin} there are exactly three good simplicial cones conjoining. Then the claim follows from \eqref{used_later_VII}. This kicks off an induction that will finish the proof. See the proof of prop.5.9 in \cite{Tizzano:2014roa} for the detail of the induction.

Finally we are left with the $G_2$'s the likes of $III,\,VI$, and they fall into the case of the right panel of fig.\ref{fig_conjoin}, which is already treated in the $Y^{1,0}$ case, and so will not be repeated here. They eventually contribute to the $G_2$'s associated with the southeast and northeast corners respectively. In conclusion, once we have treated all eight cones, we prove the factorisation formula
\bea G_2^C(z|\vec\reeb)&\sim&G_2(\frac{z}{\go_1+\go_2-\go_3}|\frac{2\go_1+\go_2}{\go_1+\go_2-\go_3},\frac{-\go_3}{\go_1+\go_2-\go_3},\frac{-1}{\go_1+\go_2-\go_3})\nn\\
&&G_2(\frac{z}{3\go_1+\go_2+\go_3}|\frac{2\go_1+\go_3}{3\go_1+\go_2+\go_3},\frac{2\go_1+\go_2}{3\go_1+\go_2+\go_3},\frac{-1}{3\go_1+\go_2+\go_3})\nn\\
&&G_2(\frac{z}{-2\go_2+\go_3}|\frac{{-\go_2}}{-2\go_2+\go_3},\frac{\go_1-\go_2+\go_3}{-2\go_2+\go_3},\frac{-1}{-2\go_2+\go_3})G_2(\frac{z}{-\go_3}|\frac{\go_2}{\go_3},\frac{\go_1+\go_2}{-\go_3},\frac{1}{\go_3}).\nn\eea
The four $G_2$'s correspond to Nekrasov partition functions of $\BB{C}^2\times_{\ep}S^1$ at southeast, northeast, northwest and southwest corner in fig.\ref{fig:cones2} respectively.

\subsection{Gluings and defects}\label{sec:gluingdefects}
In this section we make an attempt at explaining the geometric meaning of the subdivision that we used to construct the full partition function. This corresponds to gluing together a number of $S^5$'s through some surgery. This pattern has been inspired by the calculation of the five-sphere partition function. Recall that it is always possible to view a toric contact manifold as torus fibration over the base $B_{\mu}(X)$, and that the faces of $B_{\mu}(X)$ correspond to $S^3$ or lens spaces $S^3/\mathbb{Z}_p$. Also as the various tori must degenerate at the faces of the base, subdividing by adding lines introduces loci in the interior of the base where some tori degenerate. This essentially means that, when gluing together two triangles ($S^5$'s), along their common face ($S^3$ and the like), one removes from each $D\times S^3$, and glue the boundaries together with a twist that exchanges the tori. This process is an analogue of the 'symplectic cut' for the toric symplectic manifolds \cite{Lerman:1995}, see also \cite{toralactions}.

When we take the product over the cone we have to account for the now loci that appears in the interior of the cone using the prescription described in section \ref{sec:gluingcons}. The result of the products over these loci can be expressed as a ratio of special functions with less parameters indicating the fact that they account for a lower dimensional system. It is quite curious that this objects are related to the object that appears when computing BPS indices of lower dimensional field theories. This lower dimensional field theories are new degrees of freedom localized at the various loci in the interior of the base. For this reason we conjecture that our construction has an interpretation in terms of supersymmetric defects.

Recently, there has been a proposal for computing the superconformal index of the $(2,0)$ theory with defects on the five sphere \cite{Bullimore:2014upa}, see also the interesting discussion about codimension 2 defects in $\mathcal{N}=2^*$ theories in \cite{Bullimore:2014awa}. Notice that these constructions are not the same as what we did, because we never inserted supersymmetric defects in our theory, instead we understand them as emerging from the type of geometry that we consider. Nevertheless it is quite interesting that the authors of \cite{Bullimore:2014upa} found that the 5d theory on the sphere can support two kinds of defects: those with codimension 2 and with codimension 4. The contributions that they account for the defect in codimension 2 and 4 are respectively very similar to the one that we found  respectively for the propagator and the loop contributions. It would be nice to understand if there is any relationship with their construction.

\section{Conclusion}\label{s-summary}
In this paper  we constructed the full partition function for the 5D $U(1)$ $\mathcal{N}=2^*$ theory for any toric Sasaki-Einstein manifold.
 The partition function is expressed in terms of the generalised double elliptic function $G^C_2$
  associated to the moment map cone $C$ of the corresponding toric Sasaki-Einstein manifold. The construction is based on the factorisation properties of the perturbative partition function and the explicit form of the Nekrasov partition function for
   5D $U(1)$ $\mathcal{N}=2^*$ theory on $\mathbb{C}^2 \times S^1$. The full partition functions on toric Sasaki-Einstein manifolds have an
   intriguing  $SL(4,\BB{Z})$ modular property.  Moreover we propose a set of gluing rules to obtain the partition function
   by cutting the corresponding toric Sasaki-Einstein manifold into pieces with $S^5$ topology and gluing them back together.
   This way of  calculating the 5D  partition function corresponds to the insertion of defects of various codimensions. \\

   The most important result of this paper is that we have obtained the explicit form of the full partition functions for 5D
    supersymmetric gauge theories on an infinite family of  manifolds with different topology. We believe that we have observed
     only the tip of the iceberg and further study is required, especially on the structures of the partition function and the different possible
      surgeries of 5D manifolds. This may open the possibility to calculate the corresponding partition functions also on manifolds without the full toric symmetry. \\

   In the flat space $\mathbb{C}^2 \times S^1$ the 5D $U(1)$ $\mathcal{N}=2^*$ theory is the reduction of abelian 6D (2,0) theory.
    The partition function on $S^5$ can be interpreted as the superconformal index for abelian 6D (2,0) theory in
     radial quantization of $\mathbb{C}^3$. In this case $\mathbb{C}^3$ is non-singular cone over $S^5$. For a general toric Sasaki-Einstein manifold the corresponding cone is a singular Calabi-Yau cone
       (actually we do not consider the tip of  the cone as a part of geometry). It is very suggestive to think that
        our calculation is related to the index calculation of abelian 6D (2,0) theory on this Calabi-Yau cone. Although due
         to the singular nature of the cone it is not entirely clear how to perform the radial quantization. \\

   Finally let us present some highly speculative comments about the non-abelian theory. At the moment  it is not possible to
    sum up the different instanton contributions for non-abelian theories and present the result as some nice special function.
     However we believe that many structures presented in this paper will persist, especially the gluing procedure
      which involves $S^5$ pieces, in light of some results of \cite{Shabbir:2015oxa}. The geometrical manipulations should be consistent with the factorisation properties of the partition function. However in non-abelian case it is natural to expect more complicated fusion rules.

\bigskip
{\bf Acknowledgements:} We thank the anonymous referee for the comments and the suggestions on the initial draft.
   The research  is supported in part by Vetenskapsr\r{a}det
 under grant \#2014-5517, by the STINT grant and by the grant  Geometry and Physics  from the Knut and Alice Wallenberg foundation.

\appendix
\section{Special Functions}\label{app:special}

We give all necessary definitions of the special functions we employ, but we do not list all their properties and functional equations. For more details, we refer the reader to \cite{MR2101221,Tizzano:2014roa}.

The q-shifted factorial is a function defined as
\be
(z|\underline{\go})_\infty
= \prod_{j_0,\cdots,j_r=0}^{\infty}
(1- e^{2\pi i z } e^{ 2\pi i ( \underline \go \cdot \underline j )} )~.
\ee
where the infinite product converges absolutely
when $\im \omega_j > 0 \ \forall j$.
Note that we here use an abbreviation and write $(z| \underline \go )_\infty$ instead of the more common notation $(e^{2\pi i z } | e^{2\pi i \go_0},\ldots,e^{2\pi i \go_r} )_\infty$.
For other regions of parameters $\omega_i$, it is defined differently, when $\im\omega_0,\cdots,\im\omega_{k-1} <0$ and
$\im\omega_k,\cdots,\im\omega_r >0$,
 we define
\be\label{qfac}
\begin{split}
( z|\underline{\go})_\infty &=
\left\{ ( z - \go_0 - \cdots - \go_{k-1} | -\go_0, \cdots,-\go_{k-1}, \go_k, \cdots, \go_r )_\infty \right\}^{(-1)^k}\\
&= \left\{ \prod_{j_0,\cdots,j_r=0}^{\infty}
(1-e^{2\pi i z } e^{2\pi i ( -\go_0 (j_0+1) - \cdots - \go_{k-1} (j_{k-1} + 1 ) + \go_k j_k + \cdots \go_r j_r ) } ) \right\}^{(-1)^k}~.
 \end{split}
\ee
This function satisfies a variety of different functional equations, for details see \cite{MR2101221,Tizzano:2014roa}.

From the multiple q-factorial we define the \emph{multiple elliptic gamma functions} $G_r$ as
\be\label{mult_ellip_gamma}
	G_r ( z | \omega_0,\ldots,\omega_r) = \{ ( x | \underline q )_\infty  \}^{(-1)^r} ( x^{-1}q_0 q_1\cdots q_r | \underline q )_\infty~,
\ee
where $x=e^{2\pi i z}$ and $\underline q = (e^{2\pi i \go_0},\ldots,e^{2\pi i \go_r})$.
The hierarchy of $G_r$ functions include the well known theta function $\theta_0$ ($r=0$) and the ``usual'' elliptic gamma function.
The $G_r$ functions satisfy a number of nice functional equations, see \cite{MR2101221} for full details. Here we just mention a few most important ones, such as
\bea
	G_r ( z + \go_j | \underline \go ) = G_r ( z | \underline \go) G_{r-1} ( z | \go_0, \cdots,\go_{j-1},\go_{j+1},\cdots,\go_r )~,\nn
\eea
and the following nice modular property
\be \label{eq:G2modularity}
\begin{split}
G_r (z| \underline{\omega})
&= \exp \left\{
\frac{2 \pi i}{(r+2)!} B_{r+2,r+2} (z|(\underline{\omega},-1))
\right\}  \\
& \times \prod_{k=0}^r G_r
\left( \frac{z}{\omega_k} \bigg| \left(
\frac{\omega_0}{\omega_k}, \cdots,
\widecheck{\frac{\omega_k}{\omega_k}}, \cdots,
\frac{\omega_r}{\omega_k},
-\frac{1}{\omega_k}
\right) \right)~,
\end{split}
\ee
where $B_{r+2,r+2}$ is a multiple Bernoulli polynomial, defined below.

One other special function that appears in the computation of the perturbative part of the partition function is the \emph{multiple sine functions}, denoted $S_r$ and defined as
\be
	S_r ( z | \omega_1,\ldots,\omega_r) = \prod_{n\in \mathbb{Z}^r_{\geq 0} } ( z + n\cdot \underline \omega ) ( \omega_1 + \ldots + \omega_r - z + n\cdot \underline \omega )^{(-1)^{r} }~,
\ee
where the infinite product is understood as being zeta-regulated.
This hierarchy of functions are a generalization of the normal sine, which is included as the $r=1$ case, and they also satisfy multiple nice functional equations\cite{MR2101221}. For our purposes, the most important property is the factorisation property,
\be
S_r (z|\underline{\omega})
= \exp \left\{
(-1)^r \frac{\pi i}{r!} B_{r,r} (z|\underline{\omega})
\right\}
\prod_{k=1}^{r} (x_k | \underline{q_k})_\infty~,
\ee
where $\underline q_k  = (e^{2\pi i \omega_1/\omega_k} , \widecheck{\ldots, e^{2\pi i \omega_k/\omega_k}},\ldots,e^{2\pi i {\omega_r}/{\omega_k}})$.
This is of course closely related to the modular property of the $G_r$ functions above.

In these formulas, the multiple \emph{Bernoulli polynomials} show up. These are defined by the generating series
\be
	\frac{ z^r e^{zt} }{\prod_{j=1}^r ( e^{\omega_j t} - 1 ) } = \sum_{n=0}^\infty B_{r,n} ( z | \underline \omega ) \frac{t^n}{n!}
\ee
and of course also satisfies a variety of functional equations.

\subsection{Generalized multiple sine and multiple elliptic gamma functions}

For a rational convex cone $C \subset \mathbb{R}^r$ we define the generalized versions of the above functions by taking the product over all integer points inside the cone, rather than over $\mathbb{R}^r_{\geq 0}$ as above. Explicitly, for the generalized multiple sine functions we define
 \be \label{eq:S3def}
 	S_r^C ( z | \underline \omega ) = \prod_{n\in C\cap\mathbb{Z}^r } ( z + n\cdot \underline \omega ) \prod_{ n \in C^{\circ}\cap\mathbb{Z}^r } ( -z + n\cdot \underline \omega )^{(-1)^r}~,
 \ee
 where $C^\circ$ is the interior of the cone, and again the products are understood as being zeta-regulated. The mathematical details of this can be found in \cite{Tizzano:2014roa}.
 Similarly, for the generalized multiple gamma functions associated to $C$ we define them as
 \be
 	G_{r-1}^C ( z | \underline \omega ) = \prod_{n\in C\cap\mathbb{Z}^r } (1 - e^{2\pi i ( z + n\cdot \underline \omega )})^{(-1)^{r-1}} \prod_{ n \in C^{\circ}\cap\mathbb{Z}^r } (1-e^{2\pi i ( -z + n\cdot \underline \omega ) } )~.
 \ee
Comparing with the original definitions one sees restricting the second product to be over the interior of $C$ is the generalization of the shift of $\omega_1 + \ldots + \omega_r$ that appears in the usual definitions.
It is clear that if $C=\mathbb{R}^r_{\geq 0}$ these functions agree with the usual ones.
These generalized functions also enjoy some functional relations, see \cite{Tizzano:2014roa}.

Most notably, the $G_{r-1}^C$ functions have a modular property with one factor coming from each 1D face of $C$,
 \be
 	G_{r-1}^C ( z | \underline \omega ) = e^{\frac{2 \pi i}{(r+2)!} B_{r+1,r+1}^{\hat C} ( z | \underline \omega , -1) } \prod_{i=1}^{\tt n} (SK_i)^* G_{r-1} ( z | \underline \omega )~,
 \ee
 where $SK_i$ are the $SL(r+2,\mathbb{Z})$ matrices associated to the cone as described in section \ref{sec:conemodularity}, and $(SK_i)^*$ acts as a fractional linear transformation on the arguments of $G_r$, according to \eqref{eq:groupaction}.
$B_{r+1,r+1}^{\hat C}$ is the generalized Bernoulli polynomial associated to the cone $\hat C = C \times \mathbb{R}_{\geq 0}$.
This equality generalizes the modularity property of the normal $G_r$ functions shown in equation \eqref{eq:G2modularity}, and is what gives us the factorisation of the partition function.

Similarly, the generalized multiple sine has an infinite product representation with one infinite product coming from each 1d face, as 
\be
	S_r^C ( z | \underline \go ) = e^{\frac{\pi i }{r!} B_{r,r}^C ( z | \underline \go ) } \prod_{i=1}^{\tt n }  ( z | \underline \go )_\infty^{\tilde K_i} ~,
\ee
where $(z| \underline \go)_\infty^{\tilde K_i}$ is the multiple q-factorial with transformed arguments defined by
\[
 (z|\vec\reeb)_{\infty}^g=\big(\frac{z}{(g\vec\reeb)_1}\big|\frac{(g\vec\reeb)_2}{(g\vec\reeb)_1},\frac{(g\vec\reeb)_3}{(g\vec\reeb)_1}\big)_{\infty}~,~~~g\in SL(3,\BB{Z})
 \]
and where the group elements $\tilde K_i$ are defined from the cone as in equation \eqref{eq:Kidef}. Again we here abbreviate, writing $(z|a,b)_\infty$ instead of $(e^{2\pi i z } | e^{2\pi i a} , e^{2\pi i b } )_\infty$ for readability.

Appearing in these formulas are the \emph{generalized Bernoulli polynomials} defined by generalizing the definition of the usual ones, through the formal generating series
\be
	(-1)^r z^r e^{zt } \sum_{n\in C \cap \mathbb{Z}^r } e^{t (n\cdot \underline \go ) } = \sum_{n=0}^{\infty} B_{r,n}^C ( z | \underline \go ) \frac{t^n }{n!}~.
\ee
$B_{r,r}^C$ in particular is a polynomial of degree $r$ in $z$, which encodes various geometric information about the cone.

\section{Rewriting the Nekrasov partition function}\label{app:rewriting}

We start with the following expression for the instanton part of the Nekrasov partition function, as computed in \cite{Carlsson:2013jka} (see also \cite{Poghossian:2008ge} where the results are stated very clearly and explicitly):
\be
	Z^{\mathrm{inst}}_{\mathbb{C}^2\times S^1} ( m, Q, q_1,q_2) = \exp \left [ \sum_{n=1}^\infty \frac{Q^n}{n m^n } \frac{ ( m^n - q_1^n ) (m^n - q_2^n ) } { (1-Q^n)(1-q_1^n)(1-q_2^n)} \right ]~,
\ee
and then we do the following rewriting
\begin{align*}
	&Z^{\mathrm{inst}}_{\mathbb{C}^2\times S^1} ( m, Q, q_1,q_2)
	 = \exp \left [ \sum_{n=1}^\infty \frac{Q^n}{n} \frac{ m^n + m^{-n}q_1^n q_2^n - q_1^n - q_2^n } { (1-Q^n)(1-q_1^n)(1-q_2^n)} \right ] \\
	&= \exp \left [ \sum_{n=1}^\infty \frac{Q^n}{n  } ( m^n + m^{-n}q_1^n q_2^n - q_1^n - q_2^n  )  \sum_{i,j,k = 0}^\infty  (q_1^i q_2^j Q^k )^n \right ] \\
	&= \exp \left [ \sum_{n=1}^\infty \sum_{i,j,k = 0}^\infty  \frac{( m^n + m^{-n}q_1^n q_2^n - q_1^n - q_2^n  )}{n}   (q_1^i q_2^j Q^{k+1} )^n \right ] \\
	&= \exp \left [ \sum_{i,j,k = 0}^\infty - \log\frac{( 1 - m q_1^i q_2^j Q^{k+1} )( 1 - m^{-1} q_1^{i+1} q_2^{j+1} Q^{k+1})}{( 1 - q_1^{i+1} q_2^{j} Q^{k+1} )( 1 - q_1^{i} q_2^{j+1} Q^{k+1} )}\right ] \\
	& = \prod_{i,j,k = 1}^\infty \frac{ ( 1 - q_1^{i+1} q_2^{j} Q^{k+1} ) ( 1 - q_1^{i} q_2^{j+1} Q^{k+1} )   } {( 1 - m q_1^i q_2^j Q^{k+1} ) 	( 1 - m^{-1} q_1^{i+1} q_2^{j+1} Q^{k+1} ) } \\
    & =\frac{ (q_1Q|q_1, q_2, Q)_{\infty}(q_2Q|q_1,q_2,Q)_{\infty}} {(mQ|q_1, q_2, Q)_{\infty}(m^{-1}Qq_1q_2|q_1, q_2, Q)_{\infty} } \\
    &= \frac{ (q_1Q|q_1, q_2, Q)_{\infty}(q_2Q|q_1,q_2,Q)_{\infty}} {(m |q_1, q_2, Q)_{\infty}(m^{-1}Q q_1q_2|q_1, q_2, Q)_{\infty} } ( m | q_1, q_2 )_{\infty} ~ .
\end{align*}
%
%
In the final step we use properties of the multiple q-factorials to rewrite it so that the denominator can be recognized as a $G_2$-function.
Similarly the perturbative part is written as
\be
	Z^{\mathrm{pert}}_{\mathbb{C}^2\times S^1} =
	\exp \left [ \sum_{n=1}^\infty \frac{(q_1 q_2)^n }{n} \frac{m^n-1}{(1-q_1^n)(1-q_2^n)} \right ]=\frac{\big(q_1q_2|q_1,q_2)_{\infty}}{\big(mq_1q_2|q_1,q_2)_{\infty}} =
	 \frac{(q_1 q_2 | q_1, q_2)_\infty } { (m | q_1, q_2 )_\infty } \frac { (m | q_1)_\infty (m | q_2 )_\infty } {  1- m }~ . \nn
\ee
We note here that slightly different versions of this expression is found throughout the literature: this reflects an ambiguity in the calculation, that we do not know how to properly fix.
Combining the two factors, and applying properties of the q-factorials as well as recognizing the $G_2$-functions, we have that the full partition function is given by
\[
	\begin{split}
	Z^{\rm full} &= \frac{ (q_1Q|q_1, q_2, Q)_{\infty}(q_2Q|q_1,q_2,Q)_{\infty}} {G_2 ( M | \epsilon_1, \epsilon_2, \tau )}  \times\frac{(q_1 q_2 | q_1, q_2)_\infty (m | q_1)_\infty (m | q_2 )_\infty } {1-m}  \\
	&= 	\frac{ (m | q_1)_\infty (m | q_2 )_\infty } {(1 - m ) } \frac{1}{G_2 ( M | \epsilon_1, \epsilon_2, \tau) }
	\frac{ (1|Q,q_1,q_2)_\infty }{ (Q |Q,q_2)_\infty (1 |q_1 ,q_2)_\infty } ( Q q_1 q_2 | q_1,q_2,Q)_\infty (Q q_2 | q_2,Q)_\infty   (q_1 q_2 | q_1, q_2 )_\infty \\
	&= \frac{(m | q_1)_\infty (m | q_2 )_\infty }{ (Q|Q)_\infty  (q_1 | q_1 )_\infty (q_2 | q_2 )_\infty  ( 1 - m ) } \frac{G_2' ( 0 | \epsilon_1, \epsilon_2, \tau ) }{G_2 ( M | \epsilon_1, \epsilon_2, \tau) }~,
	\end{split}
\]
where $q_i = e^{2\pi i \epsilon_i}$, $m=e^{2\pi i M}$ and $Q =  e^{2\pi i \tau}$. Here, we are somewhat careful with the zero modes: up until canceling a single zero mode in the last step (i.e. in writing $G_2'$), they all cancel between the different functions, making the whole expression well-defined.
One can note here that $(Q|Q)_\infty, (q_1 | q_1)_\infty$ and $(q_2 | q_2)_\infty$ are usual $\eta$-functions, up to an overall exponential factor, and that $(m|q_i)_\infty$ similarly is kind of a shifted $\eta$-function.

%


\providecommand{\href}[2]{#2}\begingroup\raggedright

 \providecommand{\href}[2]{#2}\begingroup\raggedright\endgroup

\endgroup

\end{document}